\documentclass[10pt,journal,compsoc]{IEEEtran}
\usepackage{graphicx}
\usepackage{subfigure,wrapfig2}
\usepackage{epstopdf}
\usepackage{amssymb,amsmath,amsfonts,amsbsy,amsthm}
\usepackage{caption}
\usepackage{verbatim}
\usepackage[english]{babel}
\usepackage{color}
\usepackage{xcolor}
\usepackage{ltablex,url}
\usepackage{multirow}
\usepackage[noend]{algpseudocode}
\usepackage{algorithm}
\usepackage[T1]{fontenc}
\newtheorem{definition}{Definition}

\newtheorem{insight}{Insight}

\newcommand{\app}{{\ensuremath{\sf app}}}
\newcommand{\lib}{{\ensuremath{\sf lib}}}
\newcommand{\kernel}{{\ensuremath{\sf kernel}}}
\newcommand{\os}{{\ensuremath{\sf os}}}
\newcommand{\driver}{{\ensuremath{\sf driver}}}
\newcommand{\vul}{{\ensuremath{\sf vul}}}
\newcommand{\zd}{{\ensuremath{\sf zd}}}
\newcommand{\loc}{{\ensuremath{\sf loc}}}
\newcommand{\priv}{{\ensuremath{\sf priv}}}
\newcommand{\APP}{{\ensuremath{\sf APP}}}
\newcommand{\LIB}{{\ensuremath{\sf LIB}}}
\newcommand{\KERNEL}{{\ensuremath{\sf KERNEL}}}
\newcommand{\DRIVER}{{\ensuremath{\sf DRIVER}}}
\newcommand{\state}{{\ensuremath{\sf state}}}

\newcommand{\exploitable}{{\ensuremath{\sf exploitable}}}
\newcommand{\compromised}{{\ensuremath{\sf compromised}}}
\newcommand{\initial}{{\ensuremath{\sf initial}}}
\newcommand{\InitialCompromise}{{\ensuremath{\sf IniComp}}}
\newcommand{\Weapon}{{\ensuremath{\sf Weapon}}}

\newcommand{\con}{{\ensuremath{\sf dep\_path}}}
\newcommand{\X}{{\ensuremath{\sf X}}}
\newcommand{\A}{{\ensuremath{\cal A}}}

\newcommand{\pca}{{\ensuremath{\sf pca}}}
\newcommand{\pcos}{{\ensuremath{\sf pcos}}}
\newcommand{\cor}{{\ensuremath{\sf cor}}}
\newcommand{\OS}{{\ensuremath{\sf OS}}}
\newcommand{\capa}{{\ensuremath{\sf cap}}}
\newcommand{\browser}{{\ensuremath{\sf browser}}}

\newcommand{\ie}{\textit{e.g.}}

\newcommand{\ignore}[1]{}
\newcommand{\comb}{\textcolor{black}}

\begin{document}

\title{Quantifying Cybersecurity Effectiveness of Software Diversity \thanks{A preliminary version of the present paper appeared as \cite{XuHotSoS2018Diversity}.}}


\author{Huashan Chen, Richard B. Garcia-Lebron, Zheyuan Sun, Jin-Hee Cho,~\IEEEmembership{Senior Member,~IEEE} and Shouhuai Xu
\thanks{H. Chen, R. Garcia-Lebron and Z. Sun are with the Department of Computer Science, University of Texas at San Antonio. S. Xu is with the Department of Computer Science, University of Colorado Colorado Springs; this work was partly done when he was affiliated with University of Texas at San Antonio. J.H. Cho is with the Department of Computer Science, Virginia Tech. Correspondence: {\tt sxu@uccs.edu}}
}

\IEEEtitleabstractindextext{%
\begin{abstract}
The deployment of monoculture software stacks can cause a devastating damage even by a single exploit against a single vulnerability. Inspired by the resilience benefit of biological diversity, the concept of {\em software diversity} has been proposed in the security domain. Although it is intuitive that software diversity may enhance security, its effectiveness has not been quantitatively investigated. Currently, no theoretical or empirical study has been explored to measure the security effectiveness of network diversity. In this paper, we take a first step towards ultimately tackling the problem. We propose a systematic framework that can model and quantify the security effectiveness of network diversity. We conduct simulations to demonstrate the usefulness of the framework. In contrast to the intuitive belief, we show that diversity does {\em not} necessarily improve security from a whole-network perspective. The root cause of this phenomenon is that the degree of vulnerability in diversified software implementations plays a critical role in determining the security effectiveness of software diversity.
\end{abstract}

\begin{IEEEkeywords}
Software diversity, security quantification, security metrics, cybersecurity dynamics
\end{IEEEkeywords}}

\maketitle
\IEEEdisplaynontitleabstractindextext
\IEEEpeerreviewmaketitle

\IEEEraisesectionheading{\section{Introduction} \label{sec:introduction}}

\IEEEPARstart{S}{oftware} monoculture automatically amplifies the damage of cyber attacks because a single vulnerability in the {\em software stack} can cause the compromise of all of the computers running the same vulnerable software \cite{Geer2003,Stamp:2004:RM:971617.971650}, where {\em software stack} includes the application, library, and operating system layers. To cope with the problem, researchers have proposed the idea of diversifying the software stack \cite{zhang2001heterogeneous}. {\em Network diversity} means the software stack is diversified in a computer network.

Software diversity can be achieved by two approaches: {\em natural diversity} and {\em artificial diversity}. {\em Natural diversity} often emerges from market competition, as witnessed by the presence of different vendors for the same functionality, such as Windows versus Linux for operating systems, or Chrome versus Firefox versus Internet Explorer for browsers. {\em Artificial diversity} refers to the different versions of a functionality that are independently implemented, such as {\em N-version programming} \cite{avizienis1985n}. The concept of artificial diversity was originally introduced to enhance software reliability, but nowadays it has been adopted for achieving security purposes. This intuitive assumption seems reasonable because independent implementations of a functionality are highly unlikely to contain the same vulnerabilities.

The rule of thumb is that network diversity improves security when compared with the monoculture software stack. However, this perception has not been quantitatively validated with a scientific basis, which is necessary for justifying both the cost of artificial software diversity and the effectiveness of network diversity. In this paper, we take a first step towards ultimately tackling this problem.

In this work, we make the following contributions.
{\bf First, we quantify the security effectiveness of network diversity from a whole-network perspective}, namely viewing a network as a whole. We propose a framework for modeling attack-defense interactions in a network based on a graph-theoretic model, in which a {\em node} represents a software component or function, and an {\em arc} represents a certain relation between them and has security consequences. The framework is {\em fine-grained} because it treats individual applications, library functions, and operating system kernel functions as ``atomic'' entities. This granularity allows us to realistically model cyber attacks in a flexible manner. Moreover, the framework includes a suite of security metrics that can measure the attacker's effort, the defender's effort, and the security effectiveness of network diversity. To the best of our knowledge, this is the first framework geared towards quantifying the security effectiveness of network diversity.

{\bf Second, we conduct systematic simulations to quantify the security effectiveness of network diversity.} The findings include (and will be elaborated in Section \ref{sec:case-study}):
\begin{itemize}
\item Diversity does {\em not necessarily always} improve security from a whole-network perspective, because the security effectiveness of network diversity largely depends on the security quality of the diversified implementations.
\item The {\em independence} assumption of vulnerabilities in the diversified implementations does cause an overestimate of security effectiveness in terms of the attacker's effort. 
\item Given a fixed attack capability, increasing diversity effort can lead to a higher security as long as there are always some vulnerabilities that cannot be exploited by the attacker.
\item When diversity can improve security, enforcing diversity at multiple layers leads to higher security than enforcing diversity at a single layer. 
\item Two most effective defense strategies are (i) reducing software vulnerabilities or preventing attackers from obtaining exploits, and (ii) enforcing tight access control in host-based intrusion prevention systems (e.g., any function calls or communications that are not explicitly authorized are blocked).
\end{itemize}


\noindent{\bf Paper outline}. The rest of the paper is organized as follows. Section \ref{sec:framework} presents the proposed framework. Section \ref{sec:case-study} describes our simulation experiments and insights drawn from our experimental results. Section \ref{sec:related-work} discusses the \comb{state-of-the-art} in related work. Section \ref{sec:limitations} discusses the limitations of the present study. Section \ref{sec:conclusion} concludes the paper.

\section{Representations of Security Quantification Framework} \label{sec:framework}

\begin{figure}[!htbp]
\centering
\includegraphics[width=.46\textwidth]{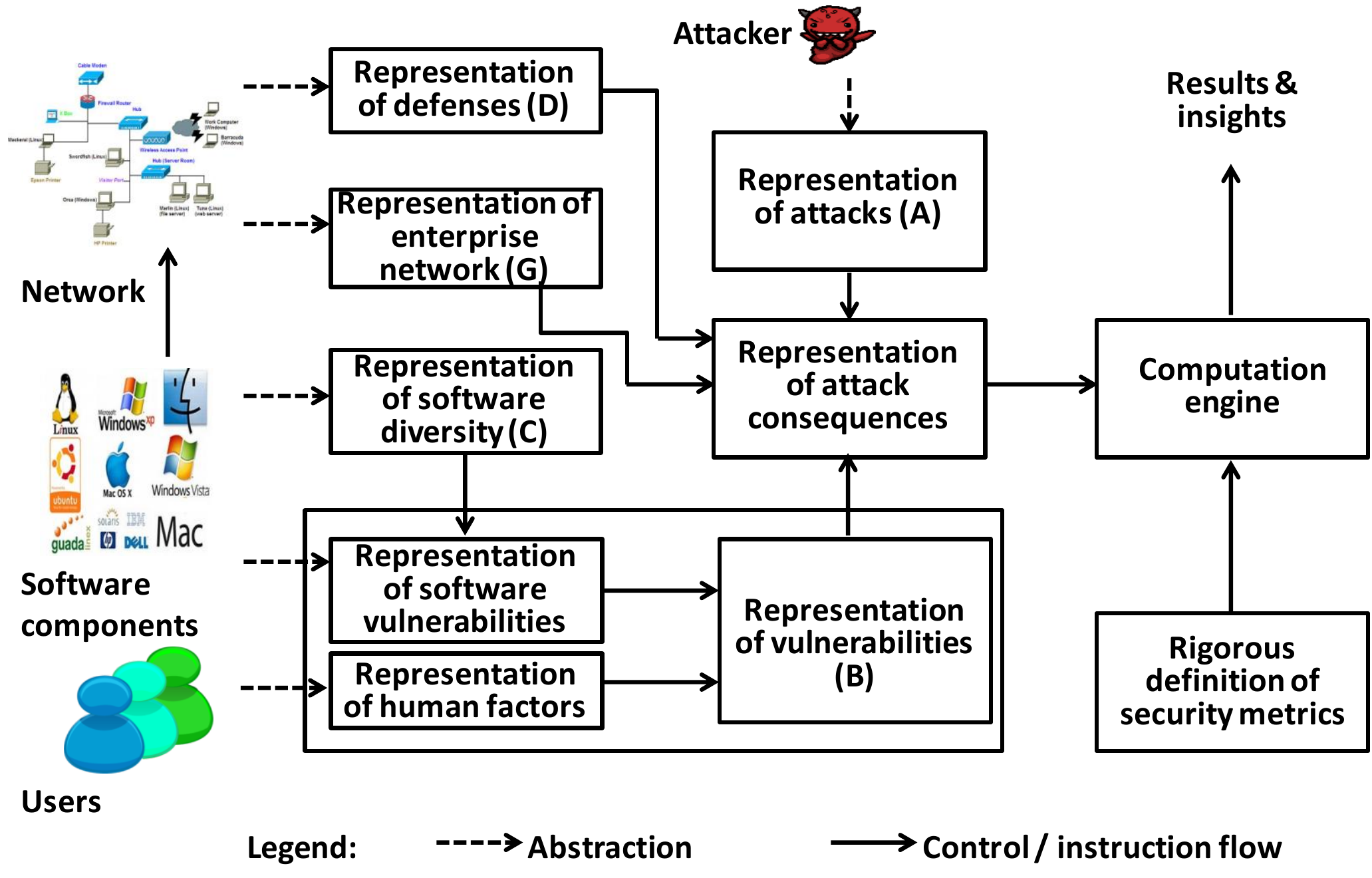}
\caption{The framework for quantifying the security effectiveness of network diversity.
\label{fig:framework}}
\end{figure}

In order to quantify the security effectiveness of network diversity, we propose \comb{a security quantification framework as described} in Fig. \ref{fig:framework}, specifying: (i) how to represent a network, its vulnerabilities, its defenses, and its software stacks; (ii) how to represent attacks; (iii) how to represent the consequences (i.e., impact) of attacks; (iv) how to define security metrics; and (v) how to compute the security effectiveness of software diversity.

At a high-level, let $G$ represent a network (e.g., an enterprise network), $A$ represent attacks against the network, $B$ represent vulnerabilities of the network's software systems and human factors, $C$ represent software stack configurations of the network, $D$ represent defenses to protect the network, and $M=\{m_i\}$ represent a set of security metrics of interest. We will discuss how $G$, $A$, $B$, $C$, $D$ and $M$ are represented later in this section. In principle, there exists a family of mathematical functions ${\cal F}_i$ for computing a network's security in terms of metric $m_i\in M$, namely
\begin{equation}
\label{eq:framework-equation}
m_i={\cal F}_i(G,A,B,C,D).
\vspace{-2mm}
\end{equation}
Intuitively, $m_i$ reflects the outcome of the interaction between attacks $A$ and defenses $D$ in a network $G$ with software stack configuration $C$ and vulnerabilities $B$. We can quantify the security effectiveness of network diversity by comparing the security \comb{levels achieved by two} software stack configurations, say $C_1$ and $C_2$, namely
$$
{\cal F}_i(G,A,B,C_1,D) ~~~\text{and}~~~ {\cal F}_i(G,A,B,C_2,D)
$$
for every ${\cal F}_i$ (i.e., for every metric $m_i\in M$ of interest).

In the following sections, we elaborate the key components of the security quantification framework.
Table \ref{table:notations} summarizes the key notations.

\begin{table}[!htbp]
\centering
\begin{tabular}{|l|p{.36\textwidth}|}
\hline
application & $\APP$ is the universe of applications; $\eta: \APP \to  \{0,1,2\}$ indicates the type of an application (client vs. server vs. other); $\app_{i,z}\in \APP$ is the $z$-th application running on computer $i$\\
\hline
library & $\LIB$ is the universe of libraries; $\lib_{i,j}\in \LIB$ is the $j$-th library running on computer $i$; $f_{i,j,z}\in \lib_{i,j}$ is the $z$-th library function in $\lib_{i,j}$\\
\hline
os         & $\OS$ is the universe of operating systems; $\os_{i}\in \OS$ is the operating system running on computer $i$; $k_{i,z}\in \os_{i}$ is the $z$-th operating system function in $\os_{i}$\\
\hline
$G_i$      & $G_i=(V_i,E_i)$ represents a computer; $V_{i}=V_{i,app} \cup V_{i,lib} \cup V_{i,os}$; $E_i=E_{i,a}\cup E_{i,l} \cup E_{i,al} \cup E_{i,lk} \cup E_{i,ak} \cup E_{i,kk}$ \\
\hline
$G=(V,E)$	& $G$ represents a network of $n$ computers, where $V=V_1\cup \ldots \cup V_n$ and $E=E_1\cup \ldots \cup E_n \cup E_0 \cup E_*$\\
\hline
diversity    & $Y^{(N)}$ is the universe of diversified implementations of $Y\in \{\APP,\LIB,\OS\}$; $N_Z$ represents the number of independent implementations of $Z$\\
\hline
vulnerability & $B$ is the universe of software vulnerabilities; $\phi(v)\subseteq B$ is the set of vulnerabilities a node $v\in V$ contains;
$\zd(\vul)$ indicates whether $\vul\in B$ is known (`0') or zero-day (`1'); $\loc(\vul)$ indicates whether $\vul$ can be exploited remotely (`1') or not ('0'); $\priv(\vul)$ indicates whether the exploitation of $\vul$ causes the attacker to obtain the {\tt root} privilege (`1') or not ('0'); $\psi(v)$ for $v\in V_i$ indicates whether the user of computer $i$ is (`1') or is not ('0') vulnerable to social engineering attacks \\
\hline
$\gamma$    & $\gamma\in [0,1]$ is the failure probability of network-based intrusion prevention mechanism; $\gamma_{(\app_1,\app_2)}\in [0,1]$ is the failure probability in blocking attacks from $\app_1$ to $\app_2$; $\gamma_{(*,\app)}\in [0,1]$ is the failure probability in blocking inbound attacks\\
\hline
$\alpha$   & $\alpha(v) \in [0,1]$ is the failure probability that a social engineering attack against $v$ is not blocked \\
\hline
$\state(v,t)$ & The probability $v\in V_{(app)}\cup V_{(os)}$ is compromised at time $t$\\
\hline
exploit       & $X$ is the set of exploits available to the attacker; $\rho(x,\vul)$ is the probability that $x\in X$ can exploit vulnerability $\vul\in B$; $\omega$ is fraction of initially compromised targets; $\capa={|\vul\in B:\exists x\in X,~\rho(x,\vul=1)|}/{|B|}$ is the fraction of software vulnerabilities that can be exploited by the attacker \\
\hline
metrics       & $M=\{m_i\}$ is a set of security metrics\\\hline
simulation    & $\zeta(v)$ is the probability that a software running at $v\in V$ or simply $v\in V$ is vulnerable; $\vartheta(\vul)$ is the probability $\vul\in B$ is remotely exploitable; $\tau(\vul)$ is the probability that $\vul\in B$ is zero-day \\\hline
\end{tabular}
\caption{Summary of notations.\label{table:notations}}
\vspace{-3mm}
\end{table}

\subsection{Networks}
\label{sec:complex-network-theoretic-description}

A network is represented by {software stacks}, {computers}, {inter-computer communication relations}, and { internal-external communication relations}.
\subsubsection{Software Stacks}
In order to represent the software stacks of computers in a network, we first identify a {\em granularity}, namely the ``atomic'' unit (e.g., treating a computer vs. a software component as an atomic object). For quantifying the security effectiveness of network diversity, treating each computer as an atomic unit is too coarse-grained. Instead, we consider three types of software running on a computer: {\em applications}, including the library functions defined by the applications; {\em libraries}, including standard library functions and non-standard library functions (e.g., system or third-party ones); {\em operating system} running in the kernel space.

\noindent{\bf Applications}. There are many kinds of applications, including client (e.g., browsers and email clients), server (e.g., web server, email server), peer-to-peer (P2P), and stand-alone (e.g., word processor). An application may include some library functions defined by the application developer. We treat each application as an atomic object, because (i) application is a natural unit of diversified implementation; (ii) an application is a privilege entity, meaning that if any part of an application is compromised, the entire application is compromised; (iii) an application can be an entry-point for an attacker to remotely penetrate into a computer (e.g., remote code execution); and (iv) attack damages are caused by the invocation of system calls (i.e., syscalls) made by applications.

Let $\APP$ denote the universe of applications. For computer $i$ in a network, we denote by $\app_{i,z}$ the $z$-th application running on the computer, where $\app_{i,z}\in \APP$.
We distinguish applications by defining the following mathematical function:
\begin{equation}
\label{eq:client-or-server-application}
\eta: \APP \to  \{0,1,2\}
\end{equation}
such that
\begin{equation*}
\eta(\app) = \begin{cases}
0 &\app\in \APP~\text{is a client \app}\\
1 &\app\in \APP~\text{is an Internet-facing server \app}\\
2 &\app\in \APP~\text{is an internal server \app}
\end{cases} \nonumber
\end{equation*}
This classification is plausible because each class may be subject to different attacks. For example, client applications (e.g., browsers and email clients) may be vulnerable to social engineering attacks, but the others may not be. An external attacker may directly compromise an Internet-facing server, but not an internal server unless the attacker already penetrated into the network.

\noindent{\bf Libraries}. We treat each library function as an atomic object \comb{because of the following:} (i) if a library function has a vulnerability, then an application that invokes this function is compromised when the vulnerability is exploited; and (ii) we need to distinguish the library functions that make system calls from those which do not. This is important because a library function that makes system calls can be leveraged to exploit a vulnerability in the operating system, but a library function that does not make any system call cannot be leveraged to exploit a vulnerable operating system. Let $\LIB$ denote the universe of libraries. For computer $i$ in a network, we denote by $\lib_{i,j}$ the $j$-th library running on the computer, and by $f_{i,j,z}$ the $z$-th library function in $\lib_{i,j}$, where $\lib_{i,j}\in \LIB$ and $f_{i,j,z} \in \lib_{i,j}$.

\noindent{\bf Operating systems}. Similar to the treatment of library functions, we treat each $\OS$ function as an atomic object. This is because an $\OS$ function may have a vulnerability, but the vulnerability can be exploited only when the $\OS$ function is {\em syscalled}. That is, we should differentiate the $\OS$ functions that are {\em syscalled} from those which are not. Let $\OS$ denote the universe of operating systems. For computer $i$ in a network, we denote by $\os_{i}$ the operating system running on the computer and denote by $k_{i,z}$ the $z$-th operating system function in $\os_{i}$, where $\os_{i}\in \OS$ and $k_{i,z}\in \os_i$.

\subsubsection{Computers}
\begin{figure}[!htbp]
\centering
\includegraphics[width=.46\textwidth]{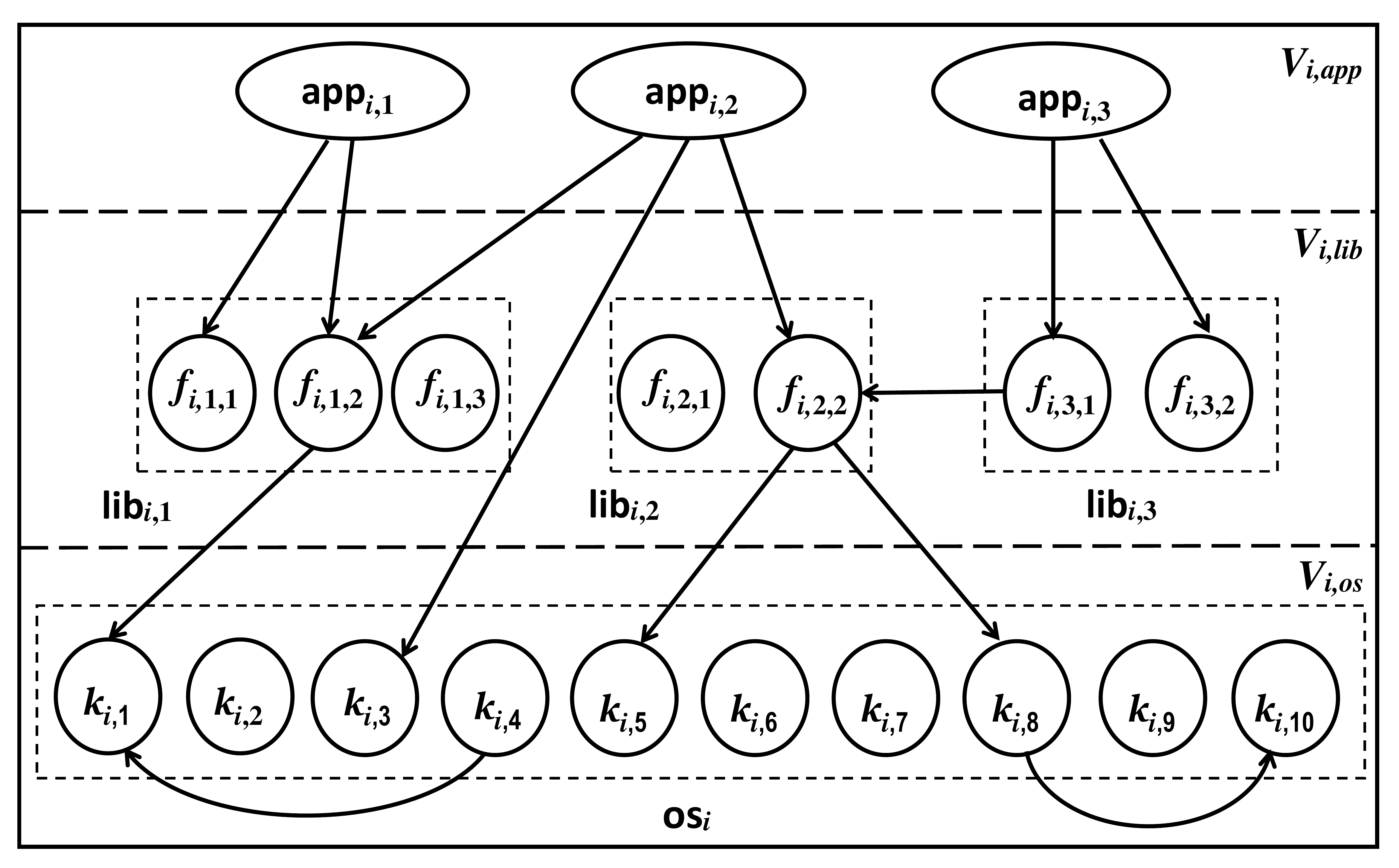}
\caption{A graph-theoretic representation of computer $i$ (or the $i$-th computer), denoted by $G_i=(V_i,E_i)$.
\label{figure:SofwareStackDependenceGraph}}
\end{figure}

Fig.~\ref{figure:SofwareStackDependenceGraph} shows a toy example of a computer, \comb{running} three applications denoted by $\app_{i,1}$, $\app_{i,2}$ and $\app_{i,3}$. Let $V_{i,app}$ denote the set of application running on computer $i$, \comb{and be defined by:}
\begin{equation}
\label{eq:Vi-app}
V_{i,app} = \{\app_{i,1},\app_{i,2},\app_{i,3}\}.
\end{equation}
There are three libraries, each of which is composed of multiple functions. For example, library $\lib_{i,1}$ consists of functions $f_{i,1,1}, f_{i,1,2},f_{i,1,3}$, and thus denoted by $\lib_{i,1}=\{f_{i,1,1}, f_{i,1,2},f_{i,1,3}\}$. Let $V_{i,lib}$ denote the library functions running on computer $i$, \comb{which is defined by:}
\begin{eqnarray}
\label{eq:Vi-lib}
V_{i,lib} &= &\lib_{i,1} \cup \lib_{i,2} \cup \lib_{i,3}\nonumber \\
&=& \{f_{i,1,1},f_{i,1,2},f_{i,1,3},f_{i,2,1},f_{i,2,2},f_{i,3,1},f_{i,3,2}\}.
\end{eqnarray}
The operating system, $\os_{i}$, has ten kernel functions, meaning $$\os_{i}=\{k_{i,1}, k_{i,2}, k_{i,3}, k_{i,4}, k_{i,5}, k_{i,6}, k_{i,7}, k_{i,8}, k_{i,9}, k_{i,10}\}.$$ Since operating system functions run at the same privilege level, we use $V_{i,os}$ to denote the operating system functions of computer $i$, \comb{and $V_{i,os}$ is given by:}
\begin{eqnarray}
\label{eq:Vi-os}
V_{i,os} = \os_{i} = \{k_{i,1},k_{i,2},k_{i,3},\ldots, k_{i,10}\}.
\end{eqnarray}

We further \comb{consider} the {\em dependence} relations between the atomic objects, namely the {\em caller-callee} relation. For example, an application may make a syscall directly or indirectly (i.e., an application calls a library function which further makes a {\em syscall}). The dependence relation should be accommodated because a vulnerability in a callee can be exploited by a caller. Fig.~\ref{figure:SofwareStackDependenceGraph} illustrates the following dependence and communication relations. Correspondingly, we model computer $i$ as a graph $G_i = (V_i,E_i)$, where $V_i$ is the {\em node set} and $E_i$ is the {\em arc set} (meaning that the graph is directed in general). \comb{$V_i$ is denoted by:}
\begin{equation}
\label{eq:Vi}
V_i=V_{i,app}\cup V_{i,lib} \cup V_{i,os},
\end{equation}
where $V_{i,app}$, $V_{i,lib}$, and $V_{i,os}$ are respectively given by Eqs. \eqref{eq:Vi-app}, \eqref{eq:Vi-lib}, \eqref{eq:Vi-os}.
The arc set \comb{is denoted by:}
\begin{equation}
\label{eq:Ei}
E_i= E_{i,al}\cup E_{i,ll}\cup E_{i,lk} \cup E_{i,ak}\cup E_{i,kk}
\end{equation}
describes the following relations:
\begin{itemize}
\item $E_{i,al}$ represents the {\em dependence} relation between applications and the library functions.
For example, in Fig.~\ref{figure:SofwareStackDependenceGraph} we have $E_{i,al}=\{(\app_{i,1},f_{i,1,1}), (\app_{i,1},f_{i,1,2}),
(\app_{i,2},f_{i,1,2}), (\app_{i,2},\\ f_{i,2,2}), (\app_{i,3},f_{i,3,1}), (\app_{i,3},f_{i,3,2})\}$.
\item $E_{i,ll}$ represents the {\em dependence} relation between library functions.
For example, in Fig.~\ref{figure:SofwareStackDependenceGraph}, we have $E_{i,l}=\{(f_{i,3,1},f_{i,2,2})\}$ because $f_{i,3,1}$ calls $f_{i,2,2}$.
\item $E_{i,lk}$ represents the {\em dependence} relation between the library functions and the operating system functions. For example, in Fig.~\ref{figure:SofwareStackDependenceGraph} we have $E_{i,lk}=\{(f_{i,1,2},k_{i,1}),(f_{i,2,2},k_{i,5})\}$.
\item $E_{i,ak}$ represents the {\em dependence} relation between applications and the operating system functions.
For example, in Fig.~\ref{figure:SofwareStackDependenceGraph}, we have $E_{i,ak}=\{(\app_{i,2},k_{i,3})\}$.
\item $E_{i,kk}$ represents the {\em dependence} relation between the operating system functions.
For example, in Fig.~\ref{figure:SofwareStackDependenceGraph} we have $E_{i,kd}=\{(k_{i,4},k_{i,1}),(k_{i,8},k_{i,10})\}$.
\end{itemize}
Putting \comb{the preceding discussion} together, we obtain the representation of computer $i$ as a graph
\begin{equation}
\label{eq:Gi}
G_i=(V_i,E_i)
\end{equation}
where $V_i$ and $E_i$ are respectively defined in Eqs. \eqref{eq:Vi} and \eqref{eq:Ei}.

\subsubsection{Inter-Computer Communication Relations Within a Network}
\begin{figure}[!htbp]
\centering
\includegraphics[width=.48\textwidth]{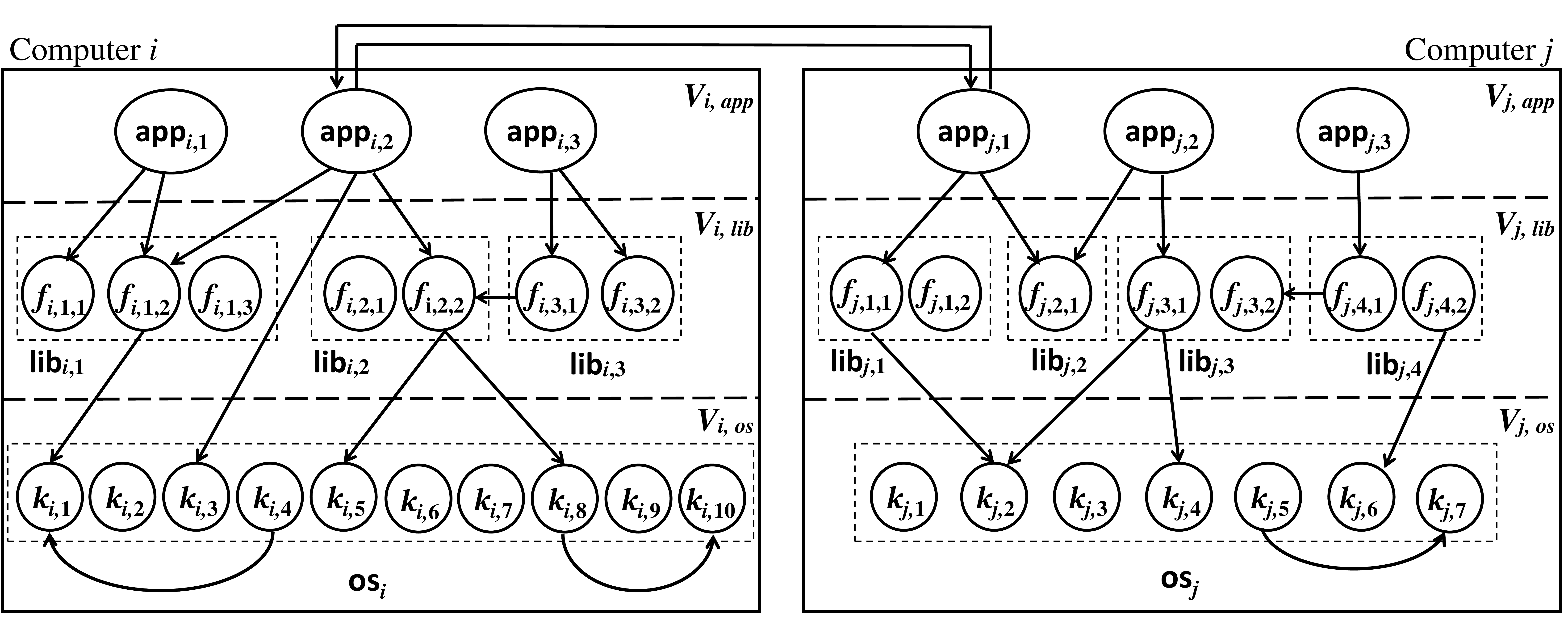}
\caption{Illustration of the {\em communication relation} with $E_0=\{(\app_{i,2},\app_{j,1})\}$.
\label{figure:network-SofwareStackDependenceGraph}}
\end{figure}

Fig.~\ref{figure:network-SofwareStackDependenceGraph} illustrates a toy network of computer $i$ and computer $j$, which are respectively described by graphs $G_i=(V_i,E_i)$ and $G_j=(V_j,E_j)$. The inter-computer {\em communication} relation describes which applications running on one computer are designed to communicate with which other applications running on another computer. \comb{We formally use} arc set $E_0$ to represent the inter-computer communication relation between applications on computer $i$ and applications on computer $j$, where
\begin{equation}
\label{eq:E0}
E_0\subseteq \{V_{i,app}\times V_{j,app}\} \cup \{V_{j,app}\times V_{i,app}\},
\end{equation}
where $1\leq i,j\leq n$ for a network of $n$ computers and $i\neq j$. In Fig.~\ref{figure:network-SofwareStackDependenceGraph}, $\app_{i,2}$ running on computer $i$ is allowed to communicate with $\app_{j,1}$ running on computer $j$ (e.g., browser to web server). Therefore, we have $E_0=\{(\app_{i,2},\app_{j,1})\}$.
Note that $e=(\app_1,\app_2)\in E_0$ does not necessarily correspond to a physical network link, be it wired or wireless. Instead, $e\in E_0$ often corresponds to a communication path.

We accommodate the communication relation because compromised computers are often used as ``stepping stones'' to attack other computers. For example, $(\app_{i,2},\app_{j,1})\in E_0$ means that the compromise of $\app_{i,2}$ may cause the compromise of $\app_{j,1}$ when $\app_{j,1}$ has a vulnerability that can be exploited remotely. In order to distinguish between two kinds of attacks that can be waged over $e\in E_0$, we partition $E_0$ into $E_{00}$ and $E_{01}$ such that $E_0=E_{00}\cup E_{01}$, where $E_{00}$ represents the attacks against clients (including peers in peer-to-peer application), and $E_{01}$ represents the attacks against servers. More specifically, we have:
\begin{itemize}
\item $(\app_{i,x},\app_{j,y})\in E_{00}$ corresponding to computers $i$ and $j$ in the network: This set represents attacks that can be launched from a server or client or peer application, say $\app_{j,y}$, against a vulnerable client application say $\app_{i,x}$ with $\eta(\app_{i,x})=0$.
\item $E_{01}=E_0\setminus E_{00}$: Any inter-computer communication other than what are accommodated by $E_{00}$.
\end{itemize}

\subsubsection{Internal-External Communication Relations}
A network is often a part of the Internet. \comb{This means} that the computers in a network may communicate with the computers outside the network. Therefore, we model the following two relations. One is the {\em internal-to-external} communication relation. A computer, say $V_i$, communicates with computers outside the network. This is done by some application running on $V_i$, say $\app_{i,1}$. We use arc set $E_{*,io}=\{(\app_{i,1},*)\}$ to describe such internal-to-external communications, where the wildcat ``$*$'' means any computer that resides outside of the network. The other is the {\em external-to-internal} communication relation. A computer, typically a server say $V_j$, is outfacing, meaning that a server application, say $\app_{j,1}$, can be accessed from any computer outside the network. We use an arc set $E_{*,oi}=\{(*,\app_{j,1})\}$ to describe such external-to-internal communications, where the wildcat ``$*$'' means that any computer outside the network can communicate with $\app_{j,1}$. Correspondingly, \comb{we define $E_*$ as:}
\begin{equation}
\label{eq:E*}
E_*=E_{*,io}\cup E_{*,oi}.
\end{equation}
Modeling $E_*$ is important because it is related to {\em initial compromise}, which deals with how an attacker penetrates into a network. For example, $E_{*,io}$ can be leveraged to wage social engineering attacks (e.g., spearfishing) and $E_{*,oi}$ can be leveraged to compromise an outfacing server.

\subsubsection{Networks}
Putting together \comb{what we have discussed}, a network of $n$ computers \comb{is represented by} $G=(V,E)$, where
\begin{equation}
\label{eq:G}
V=V_1 \cup \ldots \cup V_n~~\text{and}~~E=E_1 \cup \ldots \cup E_n \cup E_0 \cup E_*,
\end{equation}
where $G_i=(V_i,E_i)$ is given by Eq. \eqref{eq:Gi} and represents computer $i$, $E_0$ is given by Eq. \eqref{eq:E0} and represents the inter-computer communication relations between computers in the network, and $E_*$ is given by Eq. \eqref{eq:E*} and represents the internal-external communication relations.

For ease of reference, we will use $V_{(app)}$, $V_{(lib)}$ and $V_{(os)}$ to respectively denote the set of applications, libraries and operating systems running in the computers of a network, namely
\begin{eqnarray}
V_{(app)}&=&V_{1,app}\cup \ldots \cup V_{n,app}, \label{eq:Vapp}\\
V_{(lib)}&=&V_{1,lib}\cup \ldots \cup V_{n,lib}, \label{eq:Vlib}\\
V_{(os)}&=&V_{1,os}\cup \ldots \cup V_{n,os}.\label{eq:Vos}
\end{eqnarray}
We will use $v\in V$ to indicate an arbitrary node $v$.

\subsection{Network Diversity}
\label{sec:color-configuration-network-description}
\subsubsection{Software diversity}
A well-known approach to diversifying software is called {\em N-version programming} \cite{avizienis1985n}, meaning that a software has multiple independent implementations that are unlikely to have the same software bugs or vulnerabilities.
Corresponding to the representation of software stacks,
we let $\APP^{(N)}$ denote the universe of diversified implementations of the applications,
where superscript $^{(N)}$ indicates N-version programming. For application $\app\in \APP$, there is a set of independent implementations, denoted by $\app^{(N)}$, where $|\app^{(N)}|\geq 1$. Note that $|\app^{(N)}|= 1$ means that there is no diversity for this application. Similarly, we let $\LIB^{(N)}$ denote the universe of diversified implementations of the libraries, where a library $\lib \in \LIB$ has a set of independent implementations, denoted by $\lib^{(N)}$, with $|\lib^{(N)}|\geq 1$. Let $\OS^{(N)}$ denote the universe of diversified implementations of the operating systems, where \comb{an} operating system $\os\in \OS$ has a set of independent implementations, denoted by $\os^{(N)}$, with $|\os^{(N)}|\geq 1$.

\subsubsection{Network diversity} This is to diversify the software stacks of computers in networks. Consider network $G=(V,E)$ of $n$ computers, as defined in Eq. \eqref{eq:G}. The {\em software stack configuration} of computer $i$ is an assignment of specific implementations of applications, libraries, and operating system to run in computer $i$. For this purpose, we define a tuple of mathematical functions:
\begin{equation}
\label{eq:tuple-of-functions}
C=(C_{(app)},C_{(lib)},C_{(os)}),
\end{equation}
where $C_{(app)}: V_{(app)} \to \APP^{(N)}$ assigns a specific implementation of application $\app_{i,j}\in\APP^{(N)}$ to run at node $v\in V_{(app)}$, $C_{(lib)}: V_{(lib)} \to \LIB^{(N)}$ assigns a specific implementation of library $\lib_{i,j}\in\LIB^{(N)}$ to run at node $v\in V_{(lib)}$, and $C_{(os)}: V_{(os)} \to \OS^{(N)}$ assigns a specific implementation of operating system  $\os_{i}\in \OS^{(N)}$ to run at node $v\in V_{(os)}$.

\subsection{Vulnerabilities}
\label{sec:representation-of-vulnerabilities}
Following \cite{pendleton2016survey}, we consider two kinds of vulnerabilities: software vulnerabilities and human factor vulnerabilities to social engineering attacks. For a given software stack configuration of network $G=(V,E)$, we associate each node $v\in V$ with two kinds of attributes, \comb{which are} attributes associated to its software vulnerabilities and attributes associated to its user's human factor vulnerabilities.

\subsubsection{Software Vulnerabilities}
\comb{Software vulnerabilities are represented as follows.} Let $B$ denote the set of vulnerabilities that exist in the diversified implementations $\APP^{(N)}\cup\LIB^{(N)}\cup\OS^{(N)}$. When a specific implementation of $\app\in \APP^{(N)}$, $\lib\in \LIB^{(N)}$, $\os\in \OS^{(N)}$ is respectively assigned to run at some $v\in V_{(app)}$, $v\in V_{(lib)}$, $v\in V_{(os)}$, $v$ ``inherits'' or ``contains'' the software vulnerability. A vulnerability may be known to a defender or zero-day (i.e., known to the attacker but unknown to the defender).  We define a mathematical function \comb{to represent software vulnerabilities as:}
\begin{equation}
\label{eq:software-vulnerability-function}
\phi: V\to 2^B
\end{equation}
such that $\phi(v)$ represents \comb{the set of vulnerabilities contained in $v$, where} $|\phi(v)=0|$ means $v$ is not vulnerable.

We associate each vulnerability $\vul\in B$ with three attributes:
\begin{itemize}
\item $\zd$: This attribute describes whether $\vul$ is zero-day or not (i.e., an exploitation of which cannot be prevented or detected). We define predicate $\zd$ such that $\zd(\vul)=0$ means $\vul$ is known, and $\zd(\vul)=1$ means $\vul$ is zero-day.
\item $\loc$: This attribute describes whether $\vul$ can be exploited remotely or not. We define predicate $\loc$ such that $\loc(\vul)=0$ means $\vul$ cannot be exploited remotely, and $\loc(\vul)=1$ otherwise.
\item $\priv$: This attribute describes the access privileges an attacker can obtain by exploiting $\vul$.
We use predicate $\priv$ such that $\priv(\vul)=0$ means an exploitation of $\vul$ does not give the attacker the {\tt root} privilege, and $\priv(\vul)=1$ means otherwise.
\end{itemize}
These attributes allow us to accommodate attacks, such as {\em remote-2-user} attacks \cite{ghosh1999learning,poolsappasit2012dynamic,dewri2007optimal} that exploit $\vul$ with $\loc(\vul)=0$ and $\priv(\vul)=0$, {\em remote-2-root} attacks \cite{poolsappasit2012dynamic,sheyner2002automated,jha2002two} that exploit $\vul$ with $\loc(\vul)=0$ and $\priv(\vul)=1$,
and {\em user-2-root} attacks \cite{ghosh1999learning,poolsappasit2012dynamic,tavallaee2009detailed} that exploits $\vul$ with $\loc(\vul)=1$ and $\priv(\vul)=1$.
We admit that the present model does not accommodate all attacks (e.g., side-channel attacks), which will need to be accommodated in future work.

\subsubsection{Human Factor Vulnerabilities}
\comb{To represent} human factor \comb{vulnerabilities} to social engineering attacks, we define \comb{the following mathematical function:}
\begin{equation}
\label{eq:human-vulnerability-definition}
\psi: V\to \{0,1\}
\end{equation}
such that $\psi(v)$ for $v\in V_i$ indicates whether the user of computer $i$ is (`1') or is not ('0') vulnerable to social engineering attacks.

\subsection{Defenses}
\label{sec:representation-of-defenses}
In this paper, we only consider {\em preventive defenses} to prevent attacks from \comb{succeeding}, including Host-based Intrusion Prevention System (HIPS) and Network-based Intrusion Prevention System (NIPS). \comb{Since there are many defense} mechanisms that might not be feasible to model individually, we \comb{simply} model their {\em effect}. Note that the term {\em effect} is different from the term {\em effectiveness} \comb{as follows: effect includes changes introduced by a defense mechanism (e.g., cost, performance, and/or security changes), but effectiveness is limited in a positive aspect of influence (e.g., enhanced security)}. We consider two \comb{types} of preventive defenses, {\em tight} and {\em loose}, in the contexts of network-based and computer-based defenses.

For network-based preventive defenses, a {\em tight} policy is \comb{mainly related to} the enforcement of a whitelist, such that any communication attempts that are not specified by $E_0\cup E_*$ will be blocked because these communications are not deemed as \comb{necessary} by the applications. In contrast, a {\em loose} policy does not block the traffic not complying to $E_0\cup E_*$. \comb{For example, we can consider the following:} 
\begin{itemize}
\item Consider \comb{communication link} $e=(\app_1,\app_2)$, where $\app_1,\app_2\in \APP^{(N)}$ run on two different computers. If the preventive defense is {\em tight}, $e\notin E_0$ means the traffic over $e$ is blocked; otherwise, the traffic is further examined by a network-based intrusion prevention mechanism, which fails to detect an attack launched from $\app_1$ to $\app_2$ with a probability, denoted by $\gamma_{(\app_1,\app_2)}\in [0,1]$. For simplicity, we assume \comb{that} these probabilities are arc-independent, meaning $\gamma=\gamma_{(\app_1,\app_2)}$ for any $\app_1,\app_2\in \APP^{(N)}$; this corresponds to the case that the defender deploys the same network-based intrusion prevention system network-wide.

\item Consider \comb{communication link} $e=(*,\app)\in E_{*,oi}$, where $\app\in \APP^{(N)}$ and $\eta(\app)=1$. We associate $e$ with a parameter $\gamma_{(*,\app)}\in [0,1]$, which describes the probability that an inbound attack is {\em not} detected or blocked. For simplicity, we assume that these probabilities are arc-independent, meaning $\gamma=\gamma_{(*,\app)}$ for any $\app\in \APP^{(N)}$; this corresponds to the case that the same network-based intrusion detection system is used in the entire network.
\end{itemize}

For computer-based or host-based preventive defenses, a {\em tight} policy is essentially the enforcement of a whitelist, including the applications authorized to run on a computer and the list of operating system functions these applications are authorized to call. As a result, the compromise of an application does not necessarily mean the attacker can abuse the compromised application to make calls to unauthorized, but vulnerable operating system functions. Moreover, the attacker cannot run a malicious application provided by itself.
In contrast, a {\em loose} policy does not have such a whitelist. As a consequence, the compromise of an application allows the attacker to abuse the compromised application to make calls to any vulnerable operating system functions (e.g., privilege escalation). Moreover, the attacker can run any malicious application provided by itself.

In order to model computer-based or host-based preventive defenses against social engineering attacks that may be waged over $e\in E_{00}\cup E_{*,io}$, we associate each node of the following set
\begin{eqnarray}
\label{eq:nodes-subject-to-pull-based-attacks}
\{v\in V_{(app)}:\eta(v)=0\wedge ((v,*)\in E_{*,io} \vee (v,u)\in E_{00})\},
\end{eqnarray}
with parameter $\alpha_{\app}\in [0,1]$ to describe the probability that a social engineering attack against $v$ is not detected or blocked. Note that all of the applications running on computer $i$ have the same parameter $\alpha_{i}$. For simplicity, we may assume the same $\alpha$ applies to all nodes belonging to the set of Eq. \eqref{eq:nodes-subject-to-pull-based-attacks}.

\subsection{Attacks}
\label{sec:attack-description}
We describe attacks by distinguishing the {\em exploits} that are available to an attacker, and the {\em strategies} prescribing how the exploits will be used collectively. To describe the attack strategies, we define a predicate, $\state(v,t)$, for $v\in V_{(app)}\cup V_{(os)}$ such that $\state(v,t)=0$ means $v$ is not compromised at time $t$ \comb{while} $\state(v,t)=1$ means $v$ is compromised at time $t$. Note that application $v\in V_{(app)}$ can be compromised because a software vulnerability in the application or in the library function it calls is exploited, or because the underlying operating system is compromised. Note that the predicate $\state(v,t)$ is {\em not} defined for $v\in V_{(lib)}$ because library functions are loaded into the program space of an application at runtime. \comb{Now we discuss how `exploits' and `attack strategies' are represented in the proposed framework as below.}

\subsubsection{Exploits}
Let $X$ denote the set of exploits available to the attacker. We define the following mathematical function:
\begin{equation}
\label{eq:attack-success-likelihood}
\rho: X \times B \to [0,1]
\end{equation}
such that $\rho(x,\vul)$ is the success probability when applying exploit $x\in X$ against vulnerability $\vul\in B$. For simplicity, we only consider $\rho(x,\vul)=0$ or $\rho(x,\vul)=1$.

\subsubsection{Attack Strategies}
\begin{figure}
\centering
\includegraphics[width=.48\textwidth]{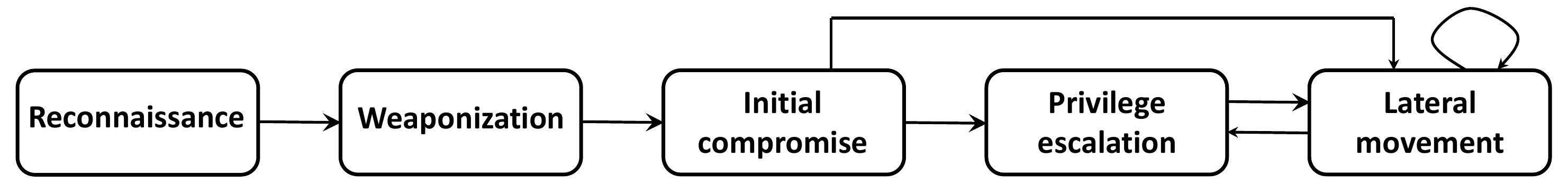}
\caption{The proposed attack lifecycle model inspired by~\cite{CyberKillChainPaper2011, mcwhorter2013apt1}.
\label{fig:attack:kill:chain}}
\vspace{-2mm}
\end{figure}

We represent attack strategies according to the attack lifecycle model highlighted in Fig. \ref{fig:attack:kill:chain}. This flexible model is adapted from the Cyber Kill Chain~\cite{CyberKillChainPaper2011} and the Attack Life Cycle~\cite{mcwhorter2013apt1}. The model has seven phases: {\em reconnaissance}, {\em weaponization}, {\em initial compromise}, {\em escalate privileges}, {\em lateral movement}, {\em persistence}, and {\em completion}. These phases are elaborated below.

\noindent{\bf Phase 1: Reconnaissance.} An attacker uses reconnaissance to collect information about a target network by identifying vulnerabilities the attacker can possibly exploit. In a given network $G=(V,E)$, we associate each node $v\in V$ with the following data structure to represent its vulnerability information:
the type $\eta(v)$ of the application running at $v\in V_{(app)}$, namely client or server; a set $\phi(v)$ of software vulnerability $v\in V$ contains; \comb{and} human factor vulnerability $\psi(v)$ of $v\in V$.
Moreover, for each vulnerability $\vul\in \phi(v)$, the attacker further knows that its attributes include the following: (i) $\zd(\vul)$, whether the vulnerability is zero-day or not; (ii) $\loc(\vul)$, whether the vulnerability can be remotely exploited or not; and (iii) $\priv(v)$, whether the exploitation of a vulnerability can cause a privilege escalation or not.

\noindent{\bf Phase 2: Weaponization.} Given $G=(V,E)$, \comb{an} outcome of the reconnaissance step, and \comb{an} attacker's set of exploits $X$, the attacker now selects some nodes $v\in V_i$ for initial compromise. For this purpose, the attacker needs to annotate the vulnerabilities that can be exploited by its exploits. There are two cases. 

In the case that $v\in V_i$ is client application $\app\in \APP^{(N)}$ running on computer $i$, meaning $\eta(v)=\eta(\app)=0$ and $(\app,*)\in E_{*,io}$, the attacker can exploit social-engineering attacks to compromise $v$ under one of the following two conditions: (i) $\app$ contains a software vulnerability, namely $\exists \vul\in \phi(v)=\phi(\app)$, or (ii) the $\app$ contains no vulnerability but a library function or operating system function that is called by the $\app$ contains a software vulnerability (i.e., there existing an access path from a secure $\app$ to a vulnerable library function or operating system function).

In order to precisely test the preceding condition (ii), there is a {\em dependence path} between two nodes $v$ and $u$ in the {\em same} computer (i.e., computer $i$ or $V_i$), if there is, according to the {\em dependence relation} defined above, a path of dependence arcs starting from node $v$ and ending at node $u$ (i.e., the software program running at node $u$ can be called, or reached, by the software program running at node $v$). We define the predicate:
\begin{equation}
\label{eq:path}
\con(v,u): V_i\times V_i \to \{\text{True},\text{False}\}
\end{equation}
such that $\con(v,u)=\text{True}$ if and only if there is a path of dependence arcs from $v$ to $u$.

\comb{To be specific}, the client application $\app$ is considered by the attacker as a candidate for {\em initial compromise} only when the following condition holds:
\begin{eqnarray} \label{eq:client-application-initial-compromisable}
(\exists \vul \in \phi(v),\exists x \in X:
\psi(v)=1 \wedge \rho(x,\vul)=1)\vee \nonumber \\
(\exists \vul \in \phi(u),\exists x \in X:(u\in V_{i,lib} \cup V_{i,os})\wedge (v\in V_{i,app})\wedge\\
\con(v,u)\wedge \psi(u)=1 \wedge \rho(x,\vul)=1).\nonumber
\end{eqnarray}
The set of client applications that can be leveraged to penetrate into a network is:
$$
\Weapon_0=\{v\in V_{i,app}: \eta(v)=0 \wedge \text{condition}~ \eqref{eq:client-application-initial-compromisable}~\text{holds}\}.
$$

In the case that $v\in V_{i,app}$ is a server application $\app\in \APP^{(N)}$ running on outfacing computer $i$, meaning $\eta(v)=\eta(\app)=1$ and $(*,\app)\in E_{*,io}$, we assume that the user of the server is not vulnerable to social engineering attacks as discussed above, meaning $\psi(v)=\psi(\app)=0$. Then, server application $\app$ can be compromised under one of the following two conditions: (i) $\app$ contains a remotely-exploitable software vulnerability, namely $\exists \vul\in \phi(v)$, where $\phi(v)=\phi(\app)$, such that $\loc(\vul)=1$; or (ii) there is a library function or operating system function that is called by $v$ (i.e., the $\app$) and that contains a remotely exploitable vulnerability.

More precisely, a server application $\app$ is considered by the attacker as a candidate for {\em initial compromise} only when the following condition holds:
\begin{eqnarray}
\label{eq:server-application-initial-compromisable}
(\exists \vul\in \phi(v),\exists x\in X:
\loc(\vul)=1\wedge \rho(x,\vul)=1) \vee \nonumber\\
(\exists \vul\in \phi(u),\exists x\in X: (u\in V_{i,lib}\cup V_{i,os}) \wedge (v\in V_{i,app}) \wedge\\
\con(v,u)\wedge \loc(\vul)=1 \wedge \rho(x,\vul)=1.\nonumber
\end{eqnarray}
The set of server applications that can be leveraged to penetrate into a network is:
$$
\Weapon_1=\{v\in V_{i,app}: \eta(v)=1 \wedge \text{condition}~ \eqref{eq:server-application-initial-compromisable}~\text{holds}\}.
$$
\comb{Summarizing the above}, the set of applications that can be leveraged to penetrate into the network is \comb{defined by:}
\begin{equation}
\label{eq:initial-compromise}
\Weapon = \Weapon_0 \cup \Weapon_1.
\end{equation}

\noindent{\bf Phase 3: Initial compromise.}
Having determined $\Weapon$ according to Eq.~\eqref{eq:initial-compromise}, the attacker will select a subset of them to penetrate into the network, according to some attack tactics. In this paper, we consider the following tactics \comb{to} reduce the chances that the attack is detected by the defender:
\begin{enumerate}
\item If the attacker can compromise the operating system by exploiting a vulnerability in $v\in V_{i,os}$, the attacker will choose to do so, even if the attacker can compromise some application belonging to $V_{i,app}$. This is because compromising the operating system causes the compromise of every $v\in V_{i,app}\cup V_{i,os}$ automatically. This tactic prevents the attacker from launching redundant attacks, and therefore possibly reduces the chance of its attacks being detected.

\item If the attacker cannot compromise the operating system on computer $i$, the attacker will compromise all of the applications on computer $i$ \comb{that can be compromised}, \comb{which is defined by:}
\begin{gather}
\{v\in V_{i,app}:v \in \Weapon \nonumber \\
\wedge (\exists x\in X, \exists \vul\in \phi(v):\rho(x,v)=1).
\end{gather}
Other tactics may be possible (e.g., the attacker may compromise some applications, depending on its objective). These attack tactics will guide the attacker to select a subset of nodes for initial compromise, denoted by:
$$
\InitialCompromise=\{v\in \Weapon: \text{attacker selects $v$ to attack} \}.
$$
\end{enumerate}

\noindent{\bf Phase 4: Privilege escalation.}
We assume the attacker wants to get the {\tt root} privilege whenever possible. Suppose the attacker only has obtained the {\tt user} privilege at a computer, meaning that the attacker has compromised some $v\in V_{i,app}$ but not $v\in V_{i,os}$. In order to escalate to the {\tt root} privilege, there are two cases, depending on the preventive defense policy is {\em tight} or {\em loose}.

In the case of {\em tight} policies, a whitelist-like mechanism is used to record the legitimate applications as well as the operating system functions they are authorized to call. Unless the host-based intrusion prevention system is compromised (i.e., the operating system is compromised), the attacker with a user privilege (by compromising an application) can neither run an arbitrary malicious program nor make any calls to unauthorized operating system functions {\em even if the latter vulnerable}. That is, a privilege escalation occurs under the following condition:
\begin{eqnarray*}
\exists v\in V_{i,app}, \exists u\in V_{i,os}, \exists \vul\in \phi(u), \exists x\in X: \\
\state(v,t)=1 \wedge \con(v,u) \wedge \rho(x,\vul)=1.
\end{eqnarray*}

In the case of {\em loose} policies, there is no whitelist-like mechanism. This means that an attacker with a user privilege (by compromising an application) can run an arbitrary malicious program or make calls to any vulnerable operating system functions to compromise them. That is, a privilege escalation occurs under the following condition:
\begin{eqnarray*}
\exists v\in V_{i,app}, \exists u\in V_{i,os}, \exists \vul\in \phi(u), \exists x\in X: \\
\state(v,t)=1 \wedge \rho(x,\vul)=1.
\end{eqnarray*}

\noindent{\bf Phase 5: Lateral movement.}
Suppose the attacker has compromised computer $i$, \comb{denoted by} $\state(v,t)=1$. Lateral movement means that the attacker attempts to compromise other computers in the network. There are two cases, depending on whether the network-based preventive defense is {\em tight} or {\em loose}.

In the case the network-based preventive defense is {\em tight}, communication over $(v,u)\notin E_0$ is blocked and therefore cannot be abused to wage attacks unless the enforcement mechanism or reference monitor is compromised (e.g., firewall). This forces the attacker to use an existing inter-computer communication relation $e\in E_{01}$ to attempt to attack another computer. Formally, a lateral movement from a compromised computer $i$ to vulnerable computer $j$ can happen under one of the following two conditions:
{\small
\begin{gather}
(\exists u\in V_{j,app},\exists \vul\in \phi(u),\exists x\in X:
\state(v,t)=1 \wedge  \nonumber \\
\state(u,t)=0 \wedge (v,u)\in E_0 \wedge \rho(x,\vul)=1 \wedge  \loc(\vul)=1) \label{eq:lateral-movement-tight-part-1}\\
\vee (\exists u\in V_{j,app},\exists w\in V_{j,lib} \cup V_{j,os}, \exists \vul\in \phi(w), \exists x\in X: \nonumber \\
(\state(v,t)=1) \wedge (\state(w,t)=0) \wedge (v,u)\in E_0 \wedge \nonumber \\
\con(u,w)\wedge \rho(x,\vul)=1 \wedge \loc(\vul)=1). \label{eq:lateral-movement-tight-part-2}
\end{gather}
}
The first condition, Eq. \eqref{eq:lateral-movement-tight-part-1}, says a vulnerable application on computer $j$ can be exploited from a compromised application on computer $i$. The second condition, Eq. \eqref{eq:lateral-movement-tight-part-2}, says a vulnerable library or operating system function on computer $j$ can be exploited from a compromised application on computer $i$.

In the case the network-based preventive defense is {\em loose}, communication over $(v,u)\notin E_0$ is {\em not} blocked by any enforcement mechanism or reference monitor and therefore can be leveraged to launch attacks. Formally, a lateral movement from a compromised computer $i$ to a vulnerable computer $j$ can happen under one of the following two conditions:
\begin{eqnarray}
(\exists u\in V_{j,app},\exists \vul\in \phi(u),\exists x\in X: \state(v,t)=1 \wedge \nonumber \\
\state(u,t)=0 \wedge \rho(x,\vul)=1 \wedge \loc(\vul)=1) \label{eq:lateral-movement-loose-part-1} \\
\vee (\exists u\in V_{j,app},\exists w\in V_{j,lib} \cup V_{j,os}, \exists \vul\in \phi(w),\exists x\in X: \nonumber\\
\state(v,t)=1 \wedge \state(w,t)=0\wedge \nonumber \\
\con(u,w)\in E_j \wedge \rho(x,\vul)=1 \wedge \loc(\vul)=1). \label{eq:lateral-movement-loose-part-2}
\end{eqnarray}
Note that Eqs. \eqref{eq:lateral-movement-loose-part-1} and \eqref{eq:lateral-movement-loose-part-2} are respectively the same as Eqs. \eqref{eq:lateral-movement-tight-part-1} and \eqref{eq:lateral-movement-tight-part-2}, except that there is no requirement for $(v,u)\in E_0$ because the network-based preventive defense is loose.

\subsection{Attack Consequences}
\label{sec:attacker_consequence_description}

The attack consequence is represented by the predicate of $\state(v)$ of $v\in V$. We can define the state of an operating system as
\begin{eqnarray*}
\state(os_i) =
\left\{
\begin{array}{ll}
1 & \exists v\in \os_i ~\text{s.t.}~\state(v)=1 \\
0 & otherwise
\end{array}
\right.
\end{eqnarray*}
Moreover, we have
\begin{eqnarray*}
\forall \app_{i,j}\in V_{i,app}: \state(os_i)=1 \Longrightarrow \state(\app_{i,j})=1.
\end{eqnarray*}

\subsection{Security Metric}
\label{sec:defining-security-metrics}

\comb{We further define security metrics to measure defender's effort, attacker's effort, and security effectiveness.}
\subsubsection{\bf Defender's Effort}
Defender's effort can be measured by the following metrics:
\begin{itemize}
\item \comb{{\em Diversity parameter} ($N$): This represents the number of implementations, $N$, each software has.} Note that it is straightforward to extend this uniform parameter $N$ into a vector $(N_{(app)},N_{(lib)},N_{(os)})$ to accommodate that different software has different numbers of implementations.
\item \comb{{\em Preventive defense effort}: This metric has two categories,} {\em tight} vs. {\em loose}, where {\em tight} means the defender needs to make extra effort in figuring out which applications have to communicate with which other applications and which applications can call which libraries or syscalls. \comb{On the other hand, {\em loose} does not require any extra effort.}
\end{itemize}

\subsubsection{\bf Attacker's Effort}
Attacker's effort can be measured by the following metrics:
\begin{itemize}
\item \comb{{\em Initial compromise effort}: It refers to the fraction $\omega$ of initial compromises the attacker makes, denoted by $\omega \times |\Weapon|$.}
\item \comb{{\em Fraction $\capa$ of vulnerabilities}: This indicates the fraction of vulnerabilities that can be exploited by the attacker, denoted by $\capa=|X|/|B|$.}
\end{itemize}

\subsubsection{\bf Security Effectiveness}
Security effectiveness can be measured by two time-dependent metrics: {\em percentage of compromised applications} (\pca) and {\em percentage of compromised operating systems} (\pcos) at time $t$, namely
\begin{eqnarray}
\pca(t)&=&{|\{v\in V_{(app)}:\state(v,t)=1\}|}/{|V_{(app)}|}, \label{eq:pca}\\
\pcos(t)&=&{|\{v\in V_{(os)}:\state(v,t)=1\}|}/{|V_{(os)}|}.\label{eq:pcos}
\end{eqnarray}
while noting that we do not consider the state of libraries.

\section{Simulation-based Case Studies}
\label{sec:case-study}

In this section, we use model-guided simulations to answer the following Research Questions (RQ):
\begin{itemize}
\item RQ1: Does natural diversity always lead to higher security? If not, when?
\item RQ2: Does artificial diversity always lead to higher security? If not, when?
\item RQ3: Does the use of natural and artificial diversity together always lead to higher security? If not, when?
\item RQ4: What are the most effective defense strategies in the presence of network diversity?
\end{itemize}

\subsection{Experimental Setup}
\label{sec:network_attack_defense_setting}

\ignore{
In order to conduct simulations, we need to specify the values of the parameters, including:
$\phi(v)$ (i.e., the set of vulnerabilities a node $v\in V$ contains), $\zd(\vul)$ (i.e., a vulnerability $\vul$ is zero-day or not), $\loc(\vul)$ (a vulnerability $\vul$ can be exploited remotely or not), and $\priv(\vul)$ (the exploitation of $\vul$ causes privilege escalation or not).
In order to accommodate a set of diversified implementations, we need to consider the randomness that can be incurred in software stack configurations and the randomness that can be contained in individual software implementations (e.g., vulnerabilities). For this purpose, we introduce the following variables: $\zeta(v)=\Pr(\phi(v)\neq \emptyset)$, namely the probability that a node $v$ is vulnerable; $\tau(\vul)=\Pr(\zd(\vul)=1)$, namely the probability that a vulnerability $vul$ is zero-day; $\vartheta(\vul)=\Pr(\loc(\vul)=1)$, namely the probability that a vulnerability $vul$ can be remotely exploited; {\color{red}and $\Pr(\priv(\vul)=1)$, namely the probability that the exploitation of $\vul$ causes privilege escalation.}
}

\subsubsection{Network Environment}
\begin{figure}[htbp!]
     \centering
     \includegraphics[width=.48\textwidth]{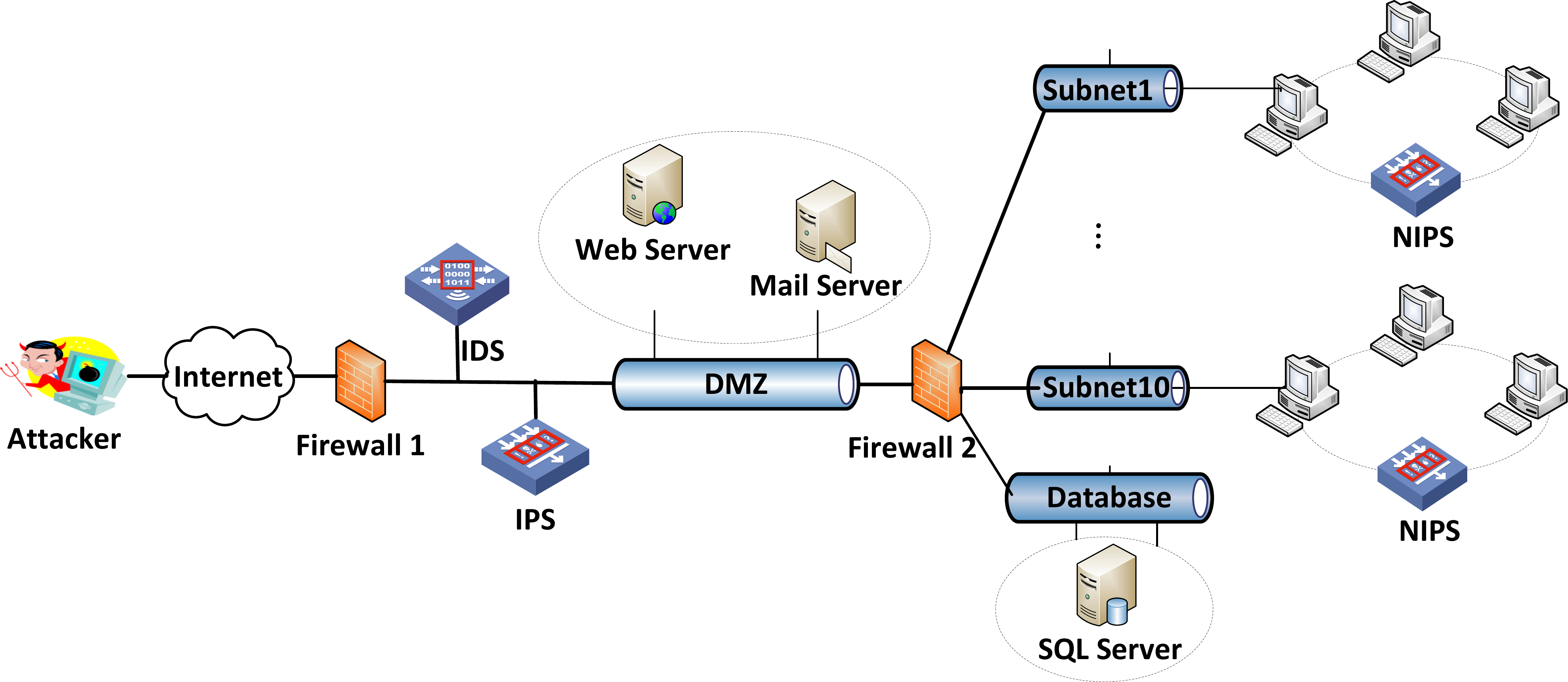}
     \caption{Example network used for the simulation study.}
     \label{fig:test bed network}
 \end{figure}

Fig. \ref{fig:test bed network} is \comb{an} example network, from which $G=(V,E)$ will be derived for the simulation study. The network has a DMZ (DeMilitarized Zone) consisting of a web server and an email server,
a database zone with a database server, and 10 subnets with each having 200 hosts. In total, the network has 2,003 computers.

For the application layer, suppose each of the three servers only runs one application (i.e., web server, email server, and database, respectively).
Suppose each computer in a subnet runs 4 applications, namely $\APP=\{$browser, email client, P2P, word processor$\}$, except for the experiment aiming to characterize the impact of the number of applications running on a computer.
For the operating system layer, suppose there are two operating systems, namely $\OS=\{\OS_1, \OS_2\}$, where $\OS_1$ offers 350 syscalls (reflecting Linux \cite{linux_syscall}) and $\OS_2$ offers 1,200 syscalls (reflecting Windows \cite{Windows_syscall}). For the library layer, suppose there are 10 libraries for $\OS_1$, including the standard library with 2,000 functions (2,000 is the number of standard libc functions in Linux \cite{libc}) and the 9 other libraries with each having 200 functions. Suppose there are 20 libraries for $\OS_2$, including the standard library with 5,000 functions (5,000 is an approximation of the standard library functions in Windows \cite{msdn}) and the 19 other libraries with each having 300 functions.

For the dependence relation in a computer, namely $E_i= E_{i,al}\cup E_{i,ll}\cup E_{i,lk}\cup E_{i,ak}\cup E_{i,kk}$, we note that $E_i$ depends on the specific software stack diversity configuration. We observe (i) precisely obtaining the $E_i$ of a given computer requires a substantial effort, and (ii) the representativeness of the given computer is always debatable. These observations suggest us to make the following simplifying assumptions.
\begin{itemize}
\item For $E_{i,al}$, we assume (i) the standard library is always called by each application, but each of the other libraries is called by each application with a $50\%$ probability; and (ii), if a library is called by an application, each function of the library is called by the application with a probability of $5\%$.

\item For $E_{i,ll}$, we assume each standard library function is called by each function in the other libraries with a probability of $5\%$.

\item For $E_{i,lk}$, we assume each operating system function is called by each standard library function with a probability of $5\%$ and is called by each function in the other libraries with a probability of $1\%$.

\item We set $E_{i,ak}=\emptyset$ because most applications will make syscalls through some libraries, rather than making syscalls directly.

\item For $E_{i,kk}$, we assume that each operating system function is called by other operating system functions with a probability of $5\%$.
\end{itemize}
For the inter-computer communication relation $E_0$, we make the following assumptions: a browser is allowed to communicate with the web server in the DMZ; an email client can connect to the email server in the DMZ to retrieve and send emails; the web server needs to communicate with the SQL server; the email clients need to communicate with each other in the enterprise network (i.e., sending emails to, and receiving emails from, each other); a P2P application needs to communicate with the other P2P applications within the same sub-network; any computer in subnet 1 can communicate with any computer in subnet 3, and any computer in subnet 2 can communicate with subnet 8; any inter-computer communication not specified above is not allowed (i.e., blocked when {\em tight} preventive defense is enforced).

For the internal-external communication relation $E_*$, we assume that a browser can access the web server outside of the network, the computers can exchange emails with the outside of the network, the P2P applications need to communicate with their peers outside of the enterprise network, the word processors can open text files received from the external network, and Internet-facing servers (i.e., web server, email server) can be accessed by external computers.

\subsubsection{Network Diversity}
For simplicity, we assume that every software has the same number $N$ of independent implementations. Given $N$ independent implementations, the tuple of configuration functions $C=(C_{(app)},C_{(lib)},C_{(os)})$ assign a specific implementation of an application, library, or operating system to run at the corresponding layer of a computer. We will consider 5 kinds of configurations that will be compared against each other: (i) $C_0$: $N=1$ (i.e., the monoculture case); (ii) $C_1$: The application, library, and operating system layers are also diversified with $N$ implementations;
(iii) $C_2$: The application layer is diversified with $N$ implementations, but the other layers are monoculture;
(iv) $C_3$: The library layer is diversified with $N$ implementations, but the other layers are monoculture; and
(v) $C_4$: The operating system layer is diversified with $N$ implementations, but the other layers are monoculture.

\ignore{

\scriptsize {
 $$C_0=(V_{(app)} \to \APP^{(1)}, V_{(lib)} \to \LIB^{(1)},V_{(os)} \to \KERNEL^{(1)}\cup \DRIVER^{(1)})$$
 $$C_1=(V_{(app)} \to \APP^{(N)}, V_{(lib)} \to \LIB^{(N)},V_{(os)} \to \KERNEL^{(N)}\cup \DRIVER^{(N)})$$
 $$C_2=(V_{(app)} \to \APP^{(N)}, V_{(lib)} \to \LIB^{(1)},V_{(os)} \to \KERNEL^{(1)}\cup \DRIVER^{(1)})$$
 $$C_3=(V_{(app)} \to \APP^{(1)}, V_{(lib)} \to \LIB^{(N)},V_{(os)} \to \KERNEL^{(1)}\cup \DRIVER^{(1)})$$
 $$C_4=(V_{(app)} \to \APP^{(1)}, V_{(lib)} \to \LIB^{(1)},V_{(os)} \to \KERNEL^{(N)}\cup \DRIVER^{(N)})$$
 }
 \normalsize

 where $C_0$ represents no diversity is enforced in the network, $C_1$ represents multiple-layer diversity is enforced in the network, $C_2$ represents only application layer diversity is enforced in the network, $C_3$ represents only library diversity is enforced in the network, $C_4$ represents only operating system layer diversity is enforced in the network.

}

\subsubsection{Vulnerabilities}
For each software belonging to $\APP^{(N)} \cup \LIB^{(N)} \cup \OS^{(N)}$, we use parameter $\zeta\in[0,1]$ to represent the probability that the software contains a vulnerability and is therefore vulnerable. For a software belonging to $\LIB^{(N)} \cup \OS^{(N)}$, the vulnerability is located at one of its functions that is chosen uniformly at random, while noting that this matter is not relevant for $\APP^{(N)}$ because an application is treated as a whole.
The attributes of vulnerability $\vul\in B$ is determined as follows. If $\vul$ is in a operating system function, then $\priv(\vul)$=1; otherwise, $\priv(\vul)$=0. We use parameter $\vartheta(\vul)$ to represent the probability that $\vul$ can be exploited remotely, namely $\Pr(\loc(\vul)=1)$. We use parameter $\tau(\vul)$ to represent the probability that $\vul$ is zero-day, namely $\Pr(\zd(\vul)=1)$.

For human factor vulnerabilities, we assume that any client computer $i$ is subject to social-engineering attacks, because it may get compromised when accessing a malicious web server or when attacked by spearfishing, namely $\psi(v)=1$ for $v\in V_i$.

\subsubsection{Defenses}
As shown in Fig. \ref{fig:test bed network}, the network uses Firewall 1 to separate the Internet from the network, uses Firewall 2 to separate the subnets from each other, and uses a NIPS to protect each subnet. We further assume that each computer runs a HIPS, which has a success probability $1-\alpha$ in detecting and blocking social-engineering attacks against computer $i$ that (i.e., its user) has a human factor vulnerability, namely $\psi(v)=1$ for $v\in V_i$ as mentioned above. For the web server and the email server in the DMZ, ports other than the specific service ports are all disabled. Firewall 2 allows the web server in the DMZ to communicate with the database server in the database zone, but block any other traffic from the DMZ to the other part of the network.

\subsubsection{Attacks}
We consider an attacker outside of the network attempting to penetrate into the network and compromise as many computers as possible. Attacks proceed according to the strategy described in Fig. \ref{fig:attack:kill:chain}.
All of the applications associated with $E_*$ can be initial compromise targets. Moreover, a compromised P2P client or email client may send a malicious message to another client to exploit the latter's vulnerability (if any).
A word processor can be exploited to spread attacks by formulating malicious payload that will be sent through either the email or the P2P application. For a given software stack diversity configuration $C=(C_{(app)},C_{(lib)},C_{(os)})$ of the software stacks, we use parameter $\capa$ to represent the fraction of vulnerabilities the attacker can exploit, where $\capa=1$ means the attacker can exploit every vulnerability.
We use parameter $\omega$ to represent the fraction of initial compromise targets, where $\omega=1$ means the attacker will initially compromise every node that is vulnerable.

\subsubsection{Simulation Algorithm}
Algorithm \ref{algorithm:vulnerablity_graph} describes the simulation algorithm, which proceeds according to the attack strategy mentioned above. The input includes $G=(V,E)$, the software stack diversity configuration $C$, the attacker's capabilities, the description of vulnerabilities $B$, and the description of defense $D$. The simulation results are presented in the $\pca$ and $\pcos$ metrics and are averaged over 100 simulation runs.

\begin{algorithm}[htbp!]
\begin{algorithmic}[1]
\Statex {\bf Input:} $G=(V,E)$ with $\eta(v)$; $A=(X,\zeta,\omega,\capa)$; $\APP^{(N)}$, $\LIB^{(N)}$, $\OS^{(N)}$; $B$ with $\zeta(\vul),\vartheta(\vul),\tau(\vul)$ for $\vul\in B$; $C$; $D=(\alpha,\gamma$,HIPS,NIPS$)$ with HIPS,NIPS$\in\{tight,loose\}$;  $T$
\Statex {\bf Output:} $\state(v,t)$ for $v\in V$ and $t=1,\ldots,T$
\State{Configure software stacks according to $C$}
\State{Assign model parameters $\alpha$ to $v$, $\gamma$ to $e\in E$, HIPS to $V_i\in V$, NIPS to $e\in E$}
\State{Simulate reconnaissance}\Comment{assuming attacker can identify all $\vul\in B$ in our simulation}
\State{Compute $\Weapon$ according to Eq. \eqref{eq:initial-compromise}}
\State{Select $\InitialCompromise$ based on $\Weapon$ and $\omega$ }
\For{$v\in V$}
\State{$\state(v,0)=0$}
\EndFor
\For{$v\in \InitialCompromise$}
\State{Simulate initial compromise}
\If{$v$ is compromised}
\State{$\state(v,1)=1$}
\EndIf
\EndFor
\For {$t \in \{2, \dots ,T\}$}
\For{each $\app\in V_{(app)}$ with $\state(v,t-1)=1$}
\State{Simulate privilege escalation and lateral movement}
\EndFor
\EndFor
\State{Return $\state(v,t)$ for $v\in V$ and $t=1,\ldots,T$ }
\end{algorithmic}
\caption{Simulation algorithm. \label{algorithm:vulnerablity_graph}}
\end{algorithm}


\subsection{RQ1: Effectiveness of Natural Diversity}
We consider natural diversity at both the application layer and the operating system layer, meaning $N=1$ for every application and operating system. Moreover, we have $N=1$
for every library.

First, we measure the security effectiveness of application-layer natural diversity by considering two browsers: $\browser_1$ and $\browser_2$ (as a simplified setting).
We consider three scenarios: (i) each computer runs $\browser_1$; (ii) each computer runs $\browser_2$; (iii) the ``hybrid'' case in which each computer runs either $\browser_1$ or $\browser_2$ with probability 0.5.
The other parameters are:
the operating system is $\OS_1$, $\gamma = 0.2$ (failure probability of NIPS), $\alpha = 0.2$ (failure probability of HIPS), $\vartheta(\vul) = 0.8$ (the probability that $\vul$ can be exploited remotely), $\tau(\vul) = 0.05$ (the probability that $\vul$ is zero-day), $\zeta(v)=0.2$ for $v\in V-\{\browser_1,\browser_2\}$ (the probability that the software running on node $v$ is vulnerable), $\capa=1$ (the worst-case scenario that the attacker has exploit for every vulnerability), $\omega=0.2$ (20\% of the nodes in $\Weapon$ are initially compromised), NIPS = $tight$, and HIPS = $tight$.

\begin{figure}[htbp!]
\vspace{-3mm}
\centering
\subfigure[Browser natural diversity]{\includegraphics[width=.23\textwidth]{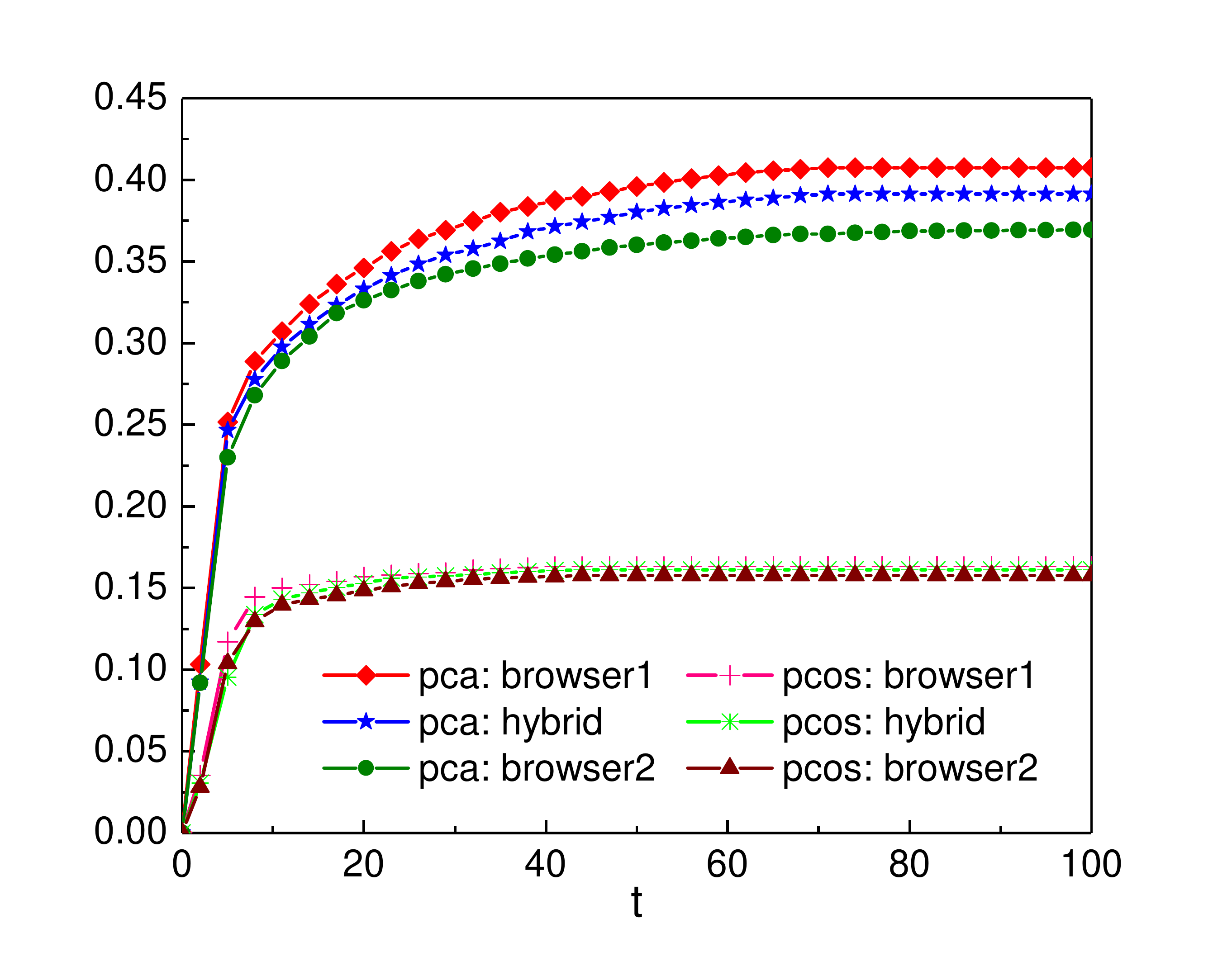}\label{fig:app-layer-natural-diversity}}
\subfigure[OS natural diversity]{\includegraphics[width=.23\textwidth]{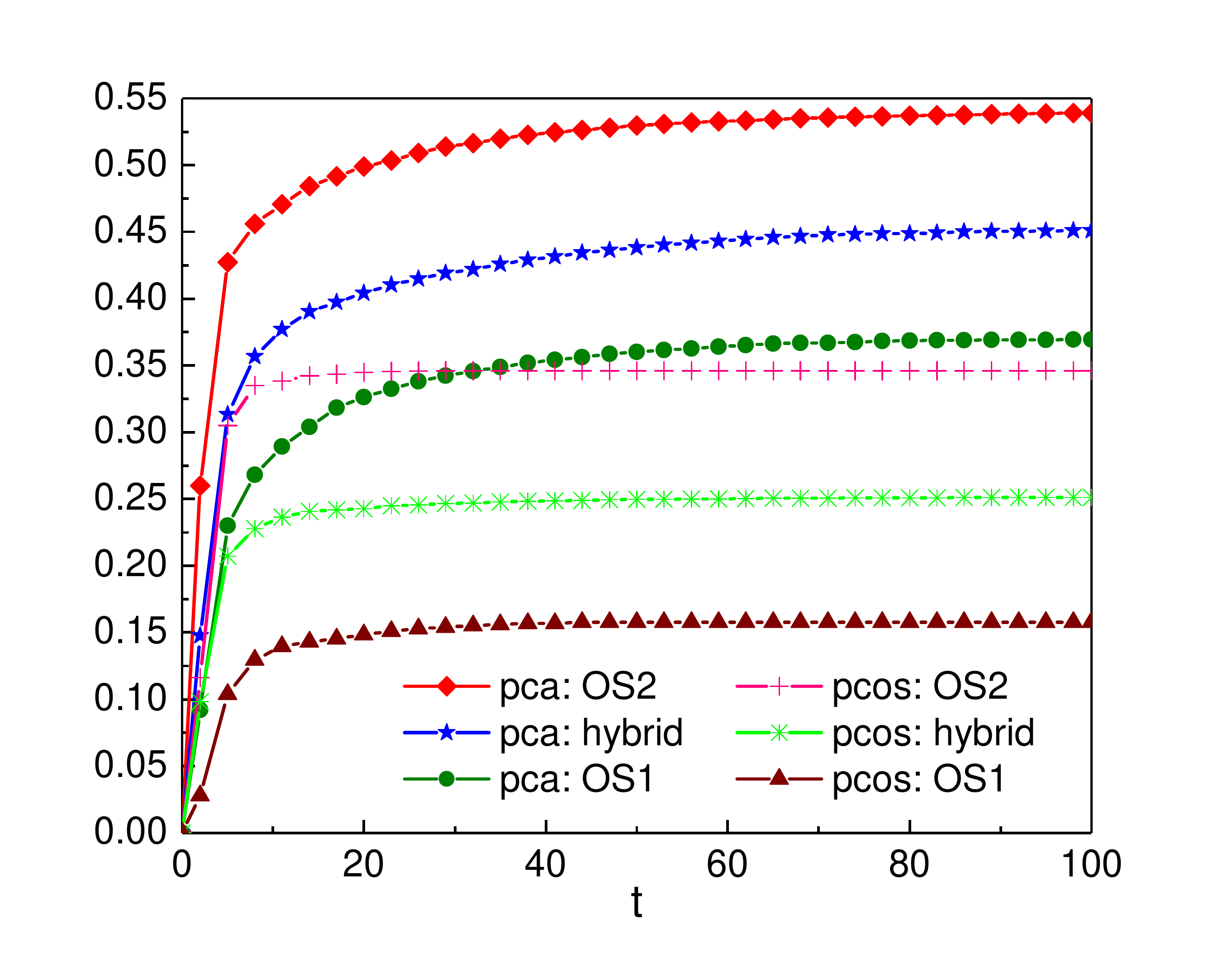}\label{fig:os-layer-natural-diversity}}
\vspace{-2mm}
\caption{Plots of $\pca(t)$ and $\pcos(t)$ with natural diversity.}
\label{mul_broswer}
\end{figure}

Fig. \ref{fig:app-layer-natural-diversity} plots $\pca(t)$ and $\pcos(t)$ with browser natural diversity, with
$\zeta(\browser_1)=0.4$ (the probability that $\browser_1$ is vulnerable) and $\zeta(\browser_2)=0.2$ (the probability that $\browser_2$ is vulnerable).
We observe that a higher (lower) browser vulnerability probability $\zeta$ leads to a higher (lower) $\pca(t)$, and the hybrid of them leads to a $\pca(t)$ somewhere in-between them.
On the other hand, $\pcos(t)$ is not affected because the underlying operating system is the same.
In addition, the percentage of compromised applications, namely $\pca(100)$, is always greater than the application vulnerable probability 0.2. This is beause an application can be compromised by exploiting a vulnerability in the application, by exploiting a vulnerability in the libraries the application invokes, or by compromising the operating system underlying it. On the other hand, the percentage of compromised operating systems, $\pcos(100)$, is always lower than the operating system vulnerable probability $0.2$. This is because
some vulnerabilities cannot be reached, and therefore cannot be exploited, by the attacker; this can happen when the HIPS enforces the $tight$ policy.

Second, we measure security effectiveness of operating system-layer natural diversity by considering the case of two operating systems: $\OS_1$ and $\OS_2$ (as a simplified setting to demonstrate the competition between various versions of Unix and Windows). We consider three scenarios: (i) every computer runs $\OS_1$; (ii) every computer runs $\OS_2$; (iii) the ``hybrid'' case in where every computer runs either $\OS_1$ or $\OS_2$ with probability 0.5.
The other parameters are:
browser being the $\browser_1$ mentioned above,
software stack diversity configuration $C_1$ with $N=1$ (which is equivalent to $C_0$ with two operating systems), $\gamma = 0.2$, $\alpha = 0.2$, $\vartheta(\vul) = 0.8$, $\tau(\vul) = 0.05$, $\zeta(v)=0.2$ for $v\in V - \{\OS_1,\OS_2\}$, $\capa=1$, NIPS= $tight$, and HIPS= $tight$.

Fig. \ref{fig:os-layer-natural-diversity} plots $\pca(t)$ and $\pcos(t)$ with operating system natural diversity, with
$\zeta(\OS_1)=0.2$ (the probability that $\OS_1$ is vulnerable) and $\zeta(\OS_2)=0.4$ (the probability that $\OS_2$ is vulnerable).
We observe that a higher (lower) operating system vulnerability probability $\zeta$ leads to a higher (lower) $\pcos(t)$, and the hybrid of them leads to a $\pcos(t)$ somewhere in between them.
When compared with the browser vulnerability probability, the operating system vulnerability probability has a more significant impact on $\pca(t)$ because a compromised operating system causes the compromise of any application running on a computer.

Summarizing the preceding discussion, we observe that when market competition leads to the emergence of a lower quality of software, security is degraded. For example, the emergence and deployment of $\OS_2$ with $\zeta=0.4$ causes $\pca(100)$ increases from 0.3694 to 0.4508 in the hybrid case, meaning a $22.04\%$ security degradation. However, if market competition leads to higher quality of software, security can be improved. For example, the emergence and deployment of
$\browser_2$ with $\zeta=0.2$ causes $\pca(100)$ decreases from 0.4076 to 0.3893 in the hybrid case, meaning a $4.49\%$ security improvement.


\begin{insight}
Natural diversity can lead to higher security as long as the diversified software implementations
have a higher security quality (i.e., containing fewer software vulnerabilities); otherwise, natural diversity can lead to lower security.
\end{insight}

\ignore{
\begin{figure}[htbp!]
\centering
\subfigure[$\pca(t)$]{\includegraphics[width=.23\textwidth]{./figures/two_os_pca.pdf}}
\subfigure[$\pcos(t)$]{\includegraphics[width=.23\textwidth]{./figures/two_os_pcos.pdf}}
\caption{Plots of $\pca(t)$ and $\pcos(t)$.}
\label{mul_layer_os}
\end{figure}
}

\subsection{RQ2: Effectiveness of Artificial Diversity}
In order to answer RQ2, we investigate a range of related sub-questions, including the impact of the dependence of vulnerabilities between diversified implementations.

\subsubsection{How does the dependence assumption of artificial diversity affect security?}
Studies \cite{knight1986experimental,eckhardt1991experimental} have showed that the independence assumption in N-version programming \comb{is} questionable because programmers tend to make the same mistakes (for example, incorrect treatment of boundary conditions). Therefore, we need to accommodate dependence between vulnerabilities. In order to describe dependence, we define the following {\em vulnerability correlation} metric.
\begin{definition}[vulnerability correlation metric]
\label{definition:cor}
Let $M$ denote the number of {\em independent} vulnerabilities in the $N$ implementations of a software program, where each vulnerability requires a different exploit. The vulnerability correlation metric, denoted by $\cor$, is defined $\cor = 1 - M/N$, where $\cor = 0$ corresponds to the extreme case that all of the $N$ vulnerabilities are independent (i.e., requiring $N$ exploits), and $\cor = (N-1)/ N$ corresponds to the other extreme case that all of the $N$ vulnerabilities can be exploited by a single exploit.
\end{definition}
Note that Definition \ref{definition:cor} implicitly assumes that each software program has at most one vulnerability. While this may be true in many cases,
the definition can be extended to accommodate the more general case of a software program contains multiple vulnerabilities.

\begin{figure}[htbp!]
\centering
\subfigure[|$\X$| vs. $\cor$]{\includegraphics[width=.23\textwidth]{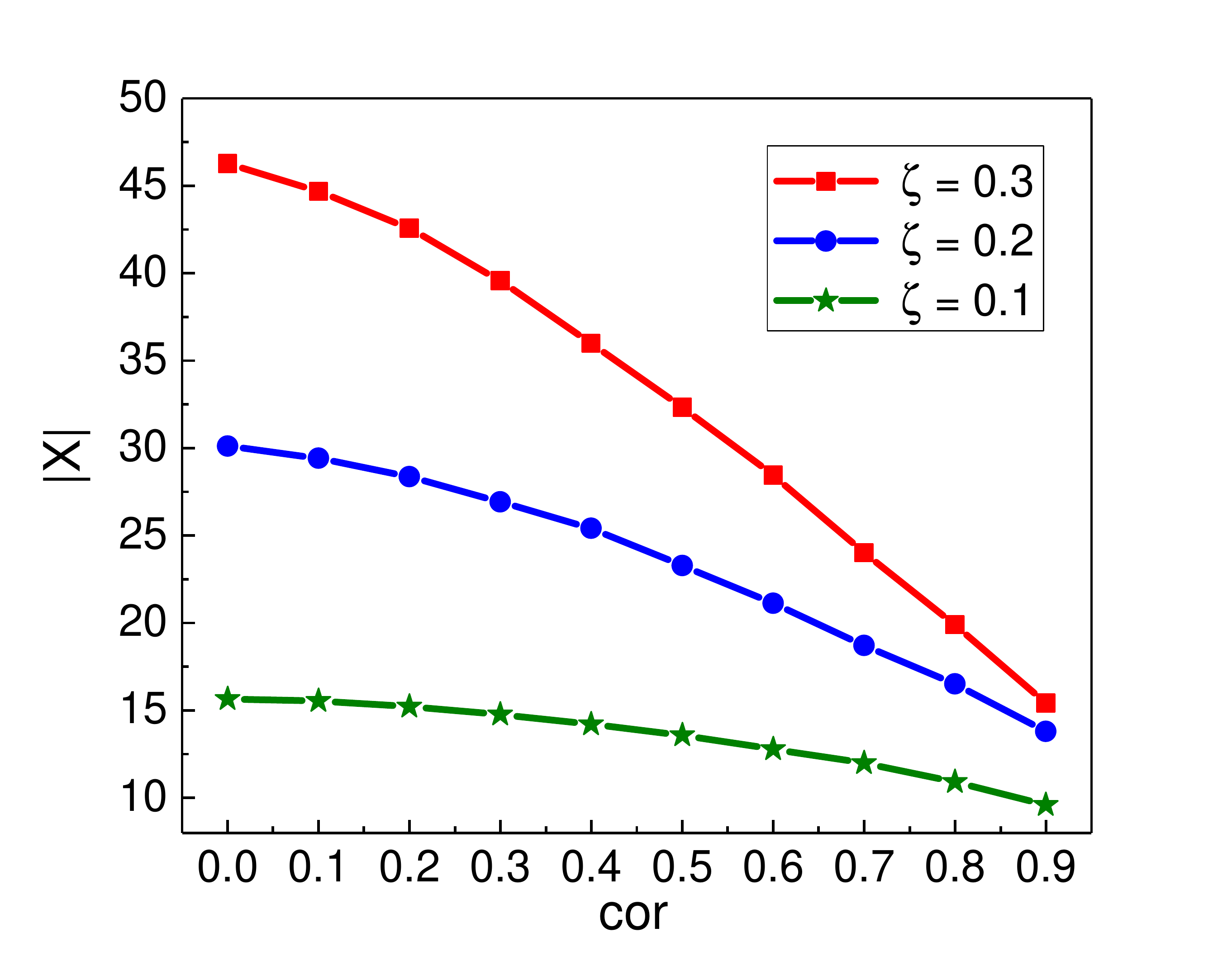}\label{fig:impact1}}
\subfigure[|$\X$| vs. $\zeta$]{\includegraphics[width=.23\textwidth]{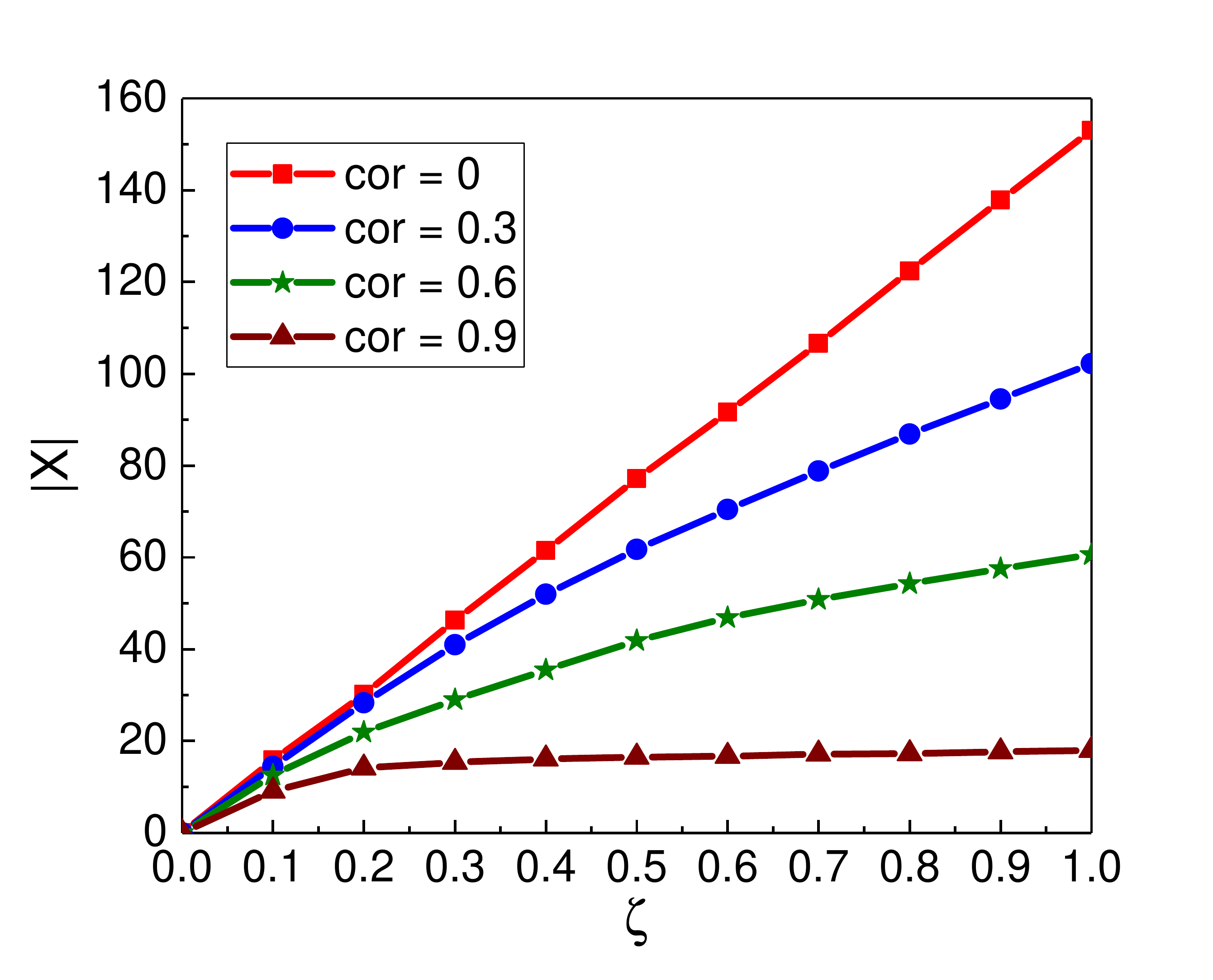}\label{fig:impact2}}
\vspace{-2mm}
\caption{Plots of attacker's effort |$\X$| with respect to fixed $\zeta$ and $\cor$.}
\label{cor-impact1}
\end{figure}

\noindent{\bf Impact of vulnerability correlation $\cor$ on attacker effort}.
In order to see the effect of vulnerability correlation $\cor$, we conduct an experiment with $N = 10$ (i.e., each software has 10 implementations), the operating system is $\OS_1$, and a fixed $\zeta$ (the probability that a software is vulnerable).

Fig. \ref{fig:impact1} shows that for a fixed $\zeta$, $|\X|$ (the number of exploits the attacker needs to obtain in order to compromise all of the vulnerable software) decreases as $\cor$ increases. Moreover, the decrease in $|\X|$ is nonlinear, and gets faster with a larger $\cor$. This confirms that dependence between vulnerabilities will reduce the security effectiveness of network diversity in terms of the attacker's effort. Furthermore, a higher software vulnerability probability $\zeta$ leads to a more substantial reduction of the attacker's effort $|\X|$ when $\cor$ increases, \comb{implying} that the attacker will benefit even more when the diversified implementations contain more vulnerabilities that are ``correlated'' with each other (i.e., lower quality).
In terms of the attacker's effort with respect to a fixed vulnerability correlation $\cor$, Fig. \ref{fig:impact2} shows that
the attacker's effort $|\X|$ grows with the software vulnerability probability $\zeta$ (indicating an increasing number of vulnerabilities). However, the growth is \comb{nonlinear} except in the case of $\cor = 0$ (i.e., the vulnerabilities are independent of each other). The stronger the vulnerability correlation (e.g., $\cor = 0.9$), the slower the increase to the attacker's effort.

\begin{insight}
The independence assumption of vulnerabilities in diversified implementations does cause an overestimate of the security effectiveness of enforcing network diversity in terms of the attacker's effort. The lower the security quality of the diversified implementations, the higher the benefit to the attacker, and the less useful the artificial diversity.
\end{insight}

\ignore{
\begin{figure}[htbp!]
\centering
\subfigure[$\pca(100)$ and $\pcos(100)$ with correlated vulnerabilities]{\includegraphics[width=.23\textwidth]{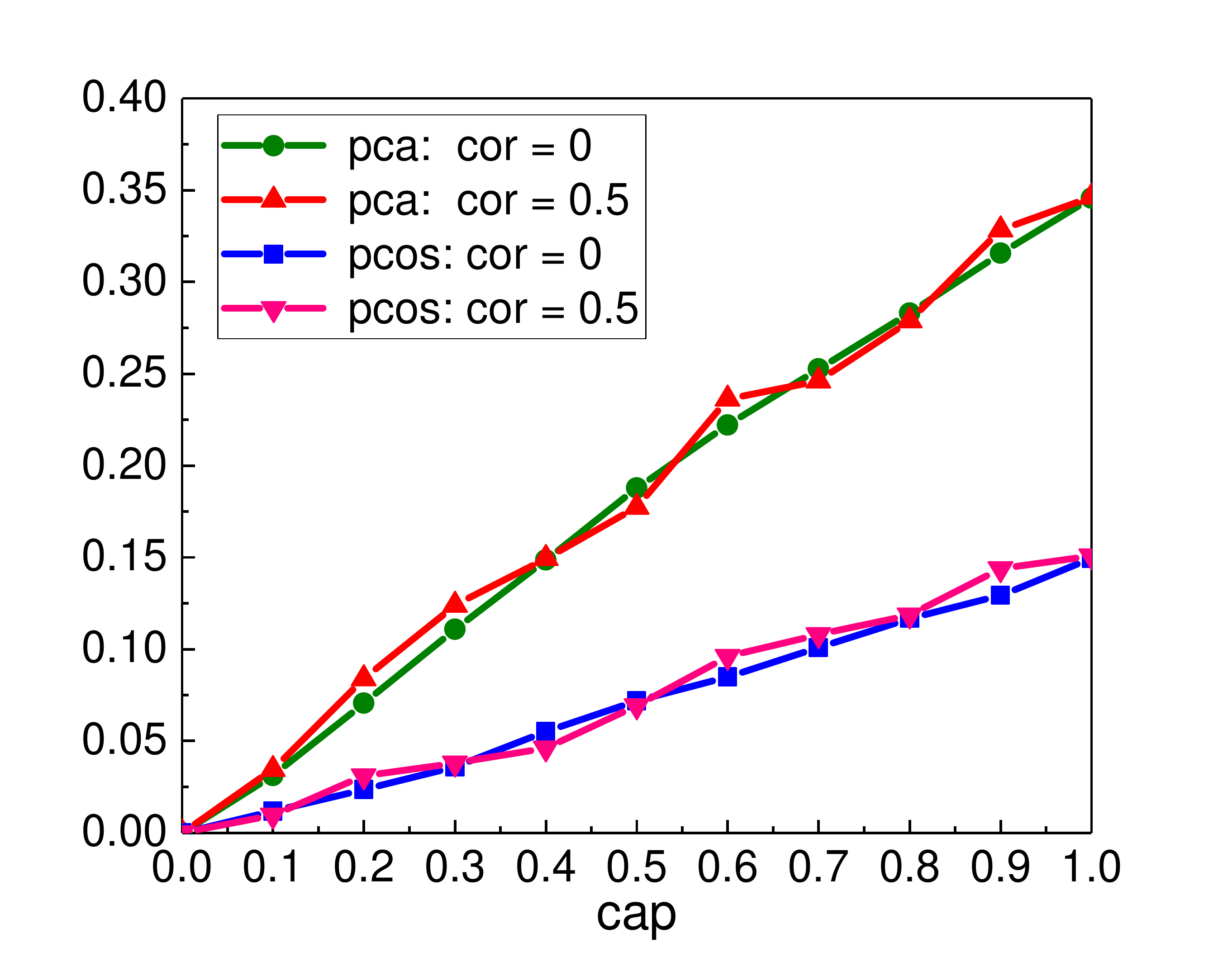}\label{col-impcat2}}
\subfigure[$\pca(100)$ and $\pcos(100)$ with varying $N$]{\includegraphics[width=.23\textwidth]{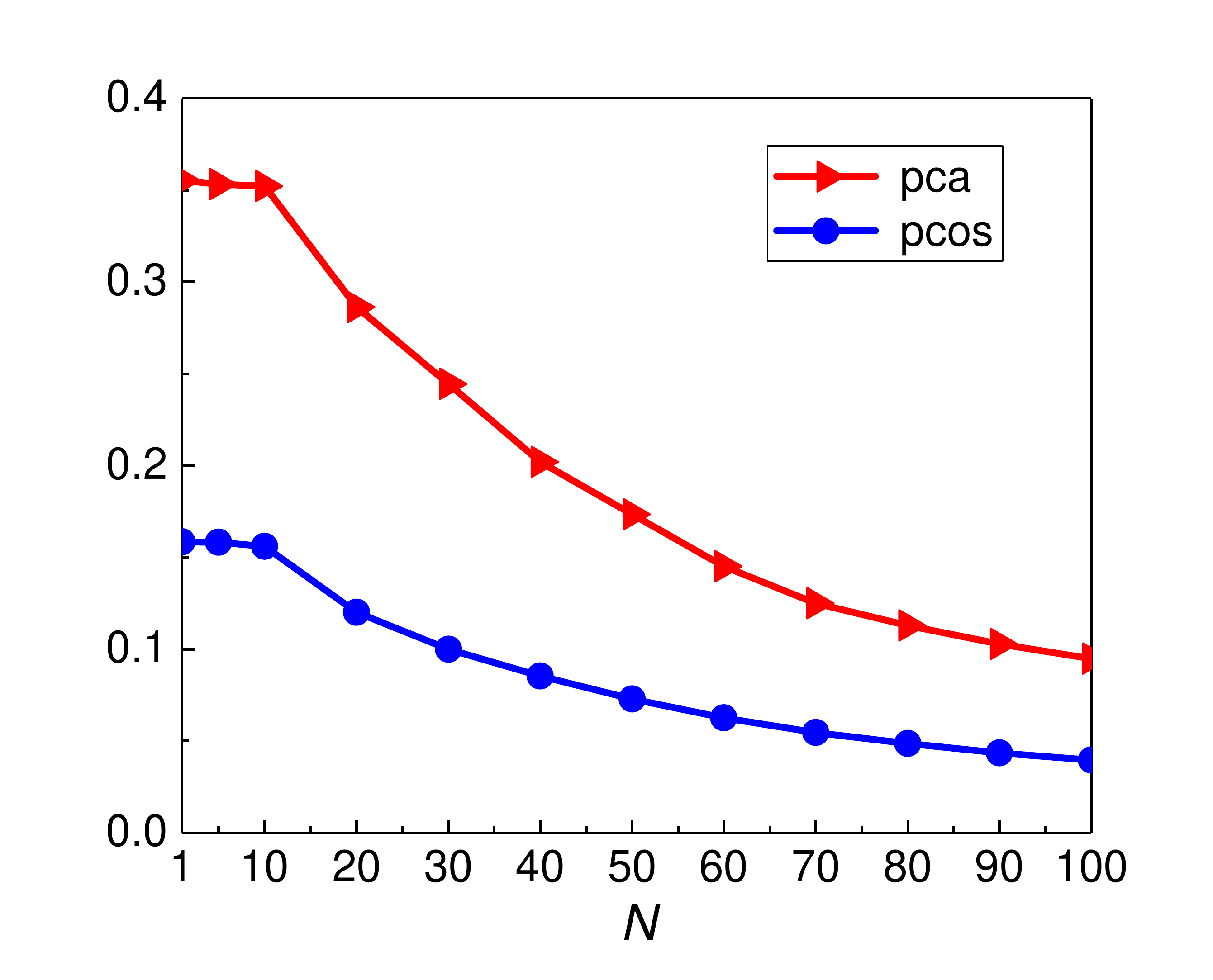}\label{versions}}
\vspace{-3mm}
\caption{Plots of $\pca(100)$ and $\pcos(100)$ with correlated vulnerabilities or varying $N$.}
\label{with-without-cor}
\end{figure}
}

\noindent{\bf Impact of vulnerability correlation $\cor$ on $\pca$ and $\pcos$}. Fig. \ref{col-impcat2} plots $\pca(100)$ and $\pcos(100)$ with the increase of attacker capability $\capa$ (the fraction of vulnerabilities for which the attacker has exploits).
\begin{wrapfigure}{r}{0.23\textwidth}
\centering
{\includegraphics[width=0.23\textwidth,height=0.21\textwidth]{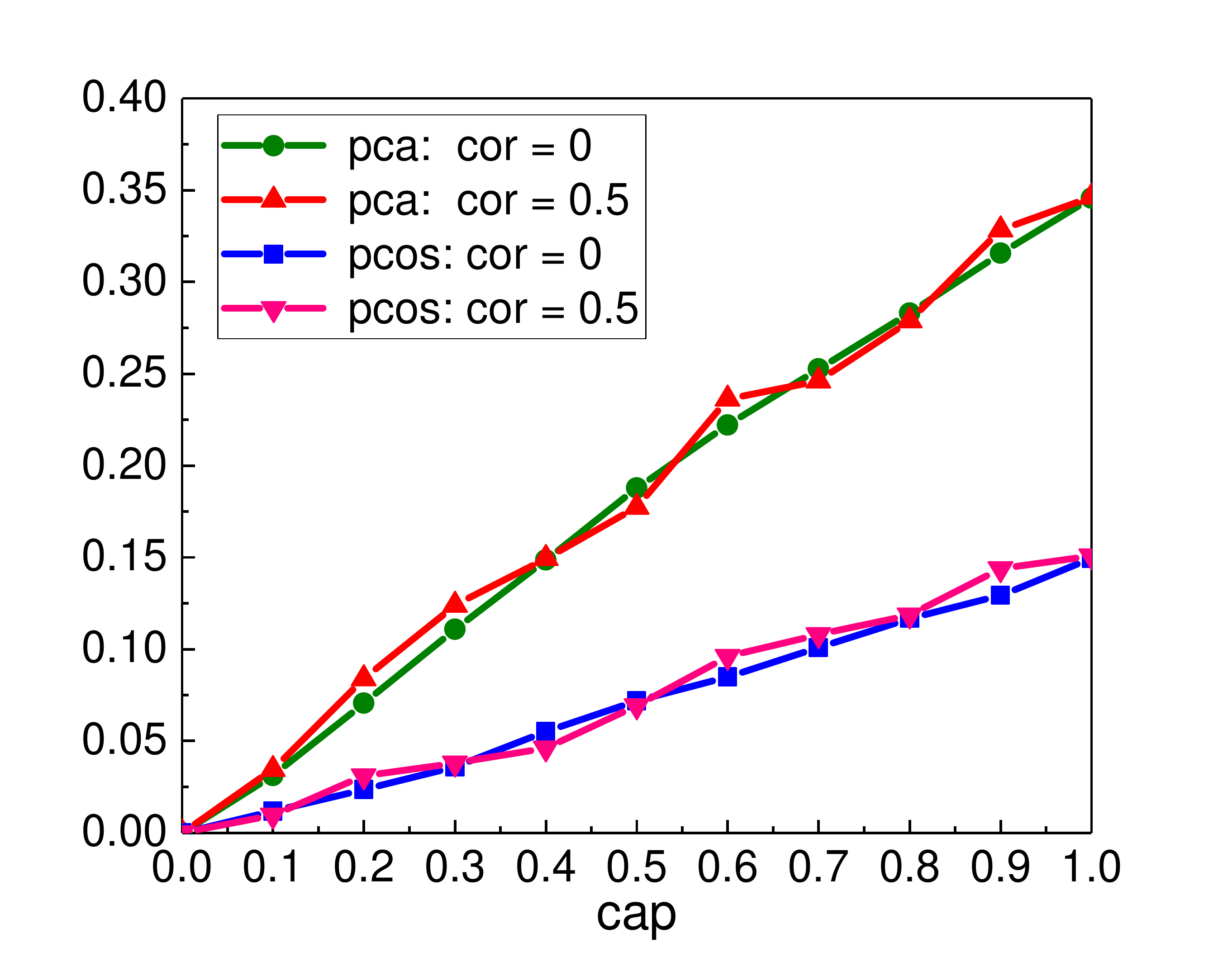}}
\caption{Plots of $\pca(100)$ and $\pcos(100)$.}
\label{col-impcat2}
\end{wrapfigure}
For a fixed vulnerability probability $\zeta = 0.2$, we compare the security consequence of $\cor = 0$ and $\cor = 0.5$. \comb{The result shows that} $\pca(100)$ with respect to $\cor = 0$ and $\pca(100)$ with respect to $\cor = 0.5$ are almost the same when $\capa = 1$ (the attacker having exploit for every vulnerability). The same phenomenon is exhibited by $\pcos(100)$ with respect to $\cor = 0$ and by $\pcos(100)$ with respect to $\cor = 0.5$. This means that vulnerability correlation $\cor$ has no effect on the security of the network against a powerful attacker because the attacker can exploit any vulnerability. On the other hand, $\cor = 0.5$ can lead to substantially higher damage in terms of $\pca$ and $\pcos$ when compared with the case of independent vulnerabilities $\cor = 0$.
\begin{insight}
The independence assumption of vulnerabilities in diversified implementations does {\em not} cause an overestimate of the security effectiveness of enforcing network diversity in terms of metrics $\pca$ and $\pcos$.
\end{insight}

\ignore{
\begin{table}[!htbp]
\tiny
\begin{tabular}{|p{1.2cm}|p{0.4cm}<{\centering}|p{0.4cm}<{\centering}|p{0.4cm}<{\centering}|p{0.4cm}<{\centering}|p{0.4cm}<{\centering}|p{0.4cm}<{\centering}|p{0.4cm}<{\centering}|p{0.4cm}<{\centering}|p{0.4cm}<{\centering}|}
\hline
\multirow{2}{*}{} & \multicolumn{9}{c|}{$\zeta$}  \\  \cline{2-10}
&  0.1      &0.2          &0.3 &0.4 &0.5 &0.6 &0.7 &0.8 &0.9\\ \hline
$C_0$ & 1.78     & 3.73 & 5.23 & 7.08 & 8.93 & 11.14 & 12.44 & 14.72 & 16.32\\ \hline
$C_1$, $\cor_2$ = 0 &  15.66   & 30.24  & 46.44 & 60.48 &  76.22 & 91.62 & 105.74 &  122.58 & 137.88          \\ \hline
$C_1$, $\cor_2$ = 0.5  & 15.175  &29.34 & 45.36 & 61.92 &  75.24  & 91.8 & 103.86  &  123.12 & 136.98            \\ \hline
$C_1$, $\cor_2$ = 1  &14.832   & 31.32 & 45.36 &61.2 &  74.34     & 92.7 & 104.94  &  122.14 & 138.14         \\ \hline
\end{tabular}
\caption{Description of
the explanatory variables
and the dependent variables $\pca$ and $\pcos$.}
\label{table:average}
\end{table}
}

\subsubsection{Does artificial diversify always lead to higher security?}
We have observed that when the attacker can exploit {\em any} vulnerability, the vulnerability correlation $\cor$ has no effect on the security with regard to $\pca$ and $\pcos$. Therefore, we need to know when artificial diversity is useful. In order to answer this question, we consider $\cor = 0$ as a representative example scenario.
We focus on $C_0$ and $C_1$ because $C_0$ in a sense reflects the state-of-the-art and $C_1$ reflects the ideal case. The other parameters are: $\APP=\{$browser, email client, P2P, word processor$\}$, operating system is $\OS_1$, $N = 10$, $\gamma = 0.2$, $\alpha = 0.2$, $\vartheta(\vul) = 0.8$, $\tau(\vul) = 0.05$, $\omega=0.2$, NIPS= $tight$, HIPS= $tight$.

\begin{figure}[htbp!]
\centering
\subfigure[Configuration $C_0$, $\zeta=0.2$]{\label{fig:subfig:a}\includegraphics[width=0.23\textwidth]{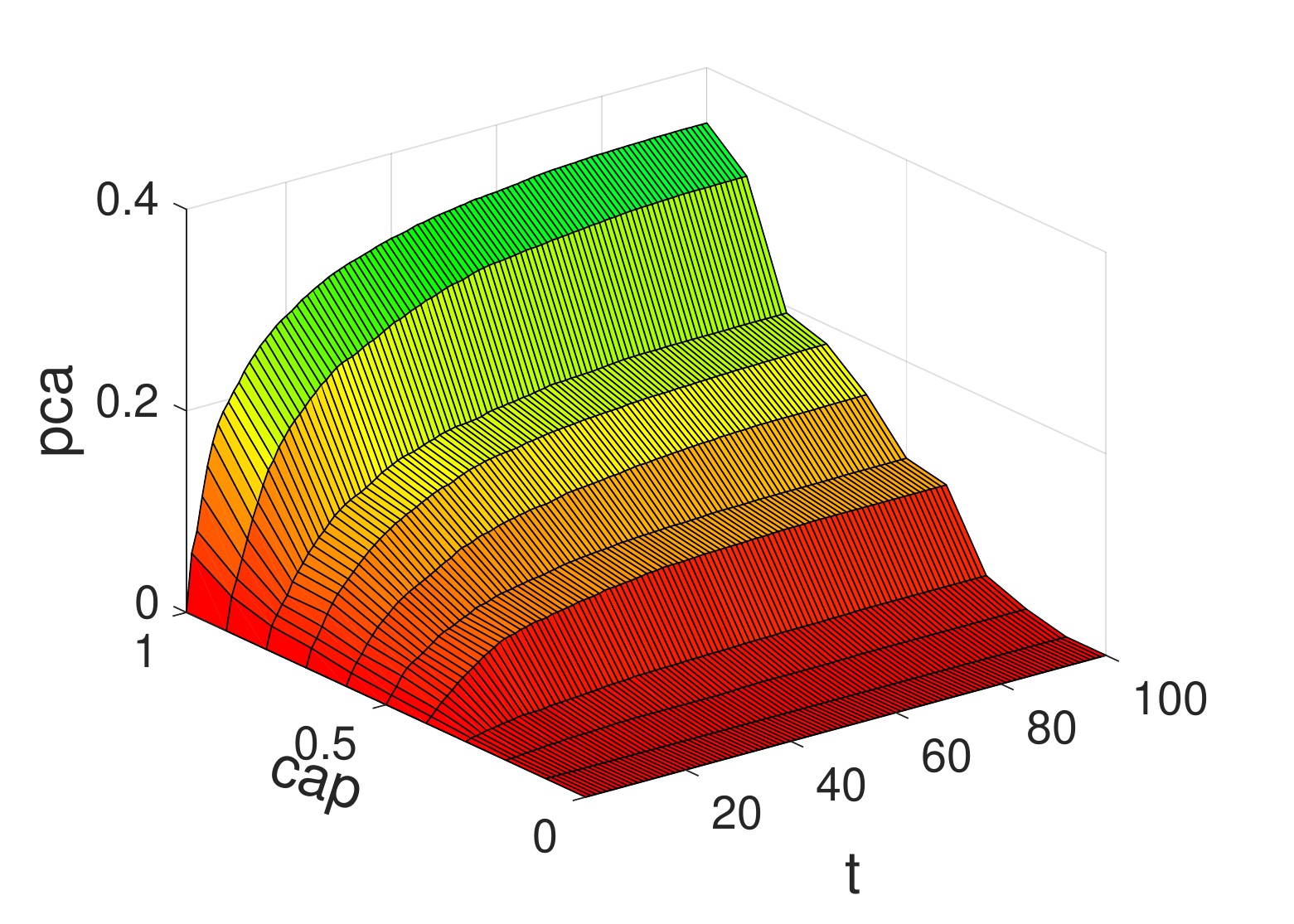}}
\subfigure[Configuration $C_1$, $\zeta=0.2$]{\label{fig:subfig:b}\includegraphics[width=0.23\textwidth]{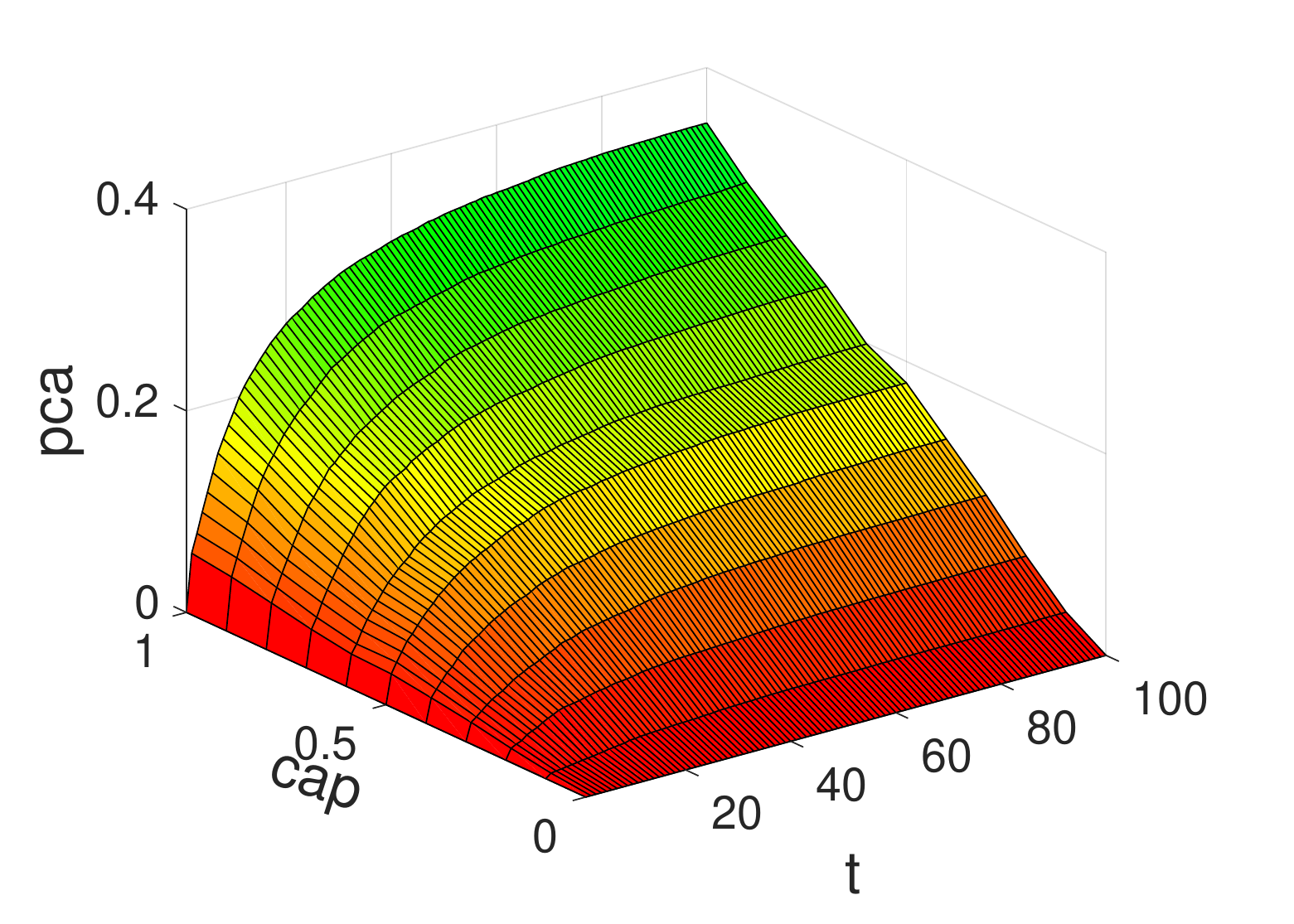}}
\subfigure[Configuration $C_0$, $\zeta=0.2$]{\label{fig:subfig:c}\includegraphics[width=0.24\textwidth]{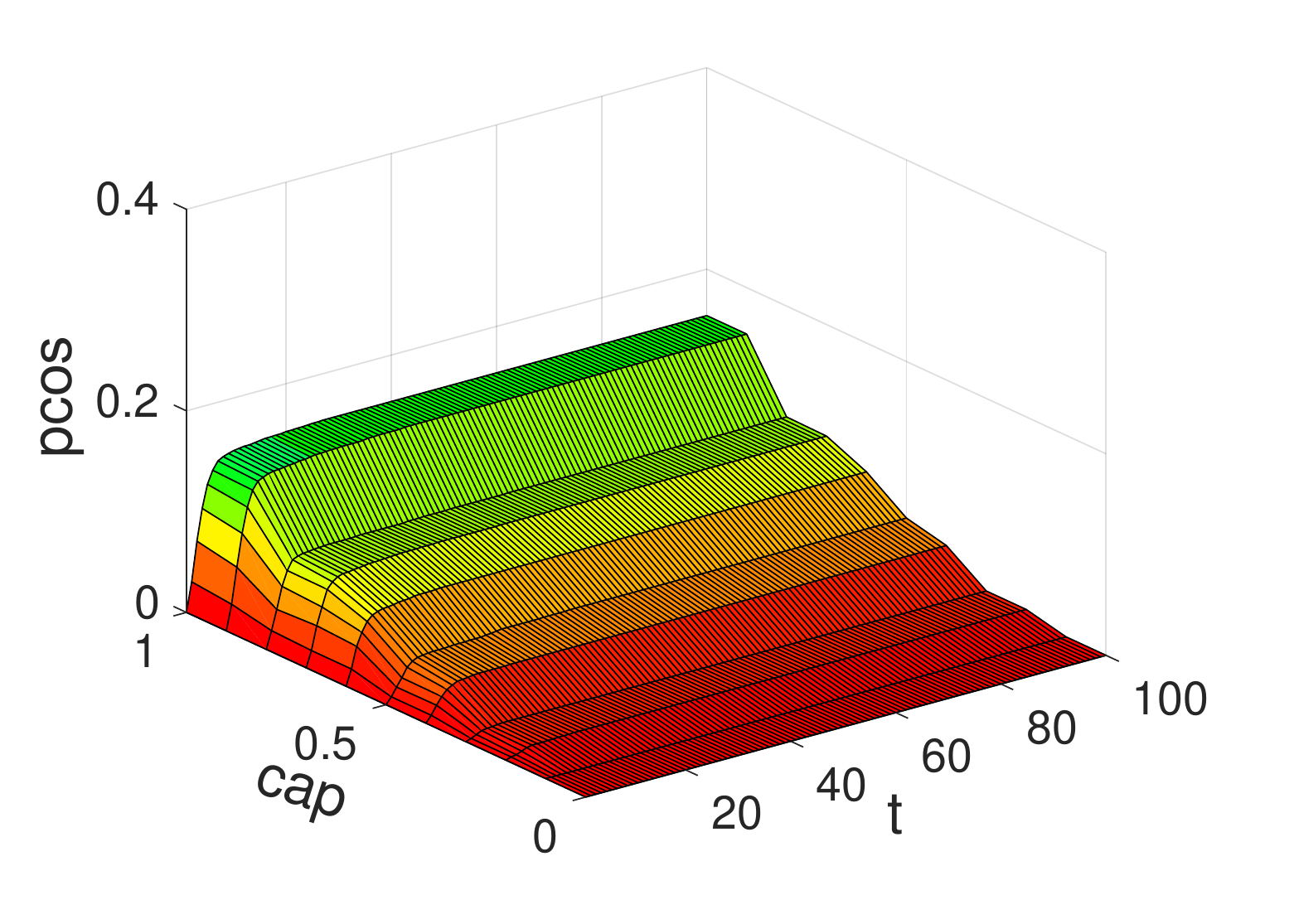}}
\subfigure[Configuration $C_1$, $\zeta=0.2$]{\label{fig:subfig:d}\includegraphics[width=0.23\textwidth]{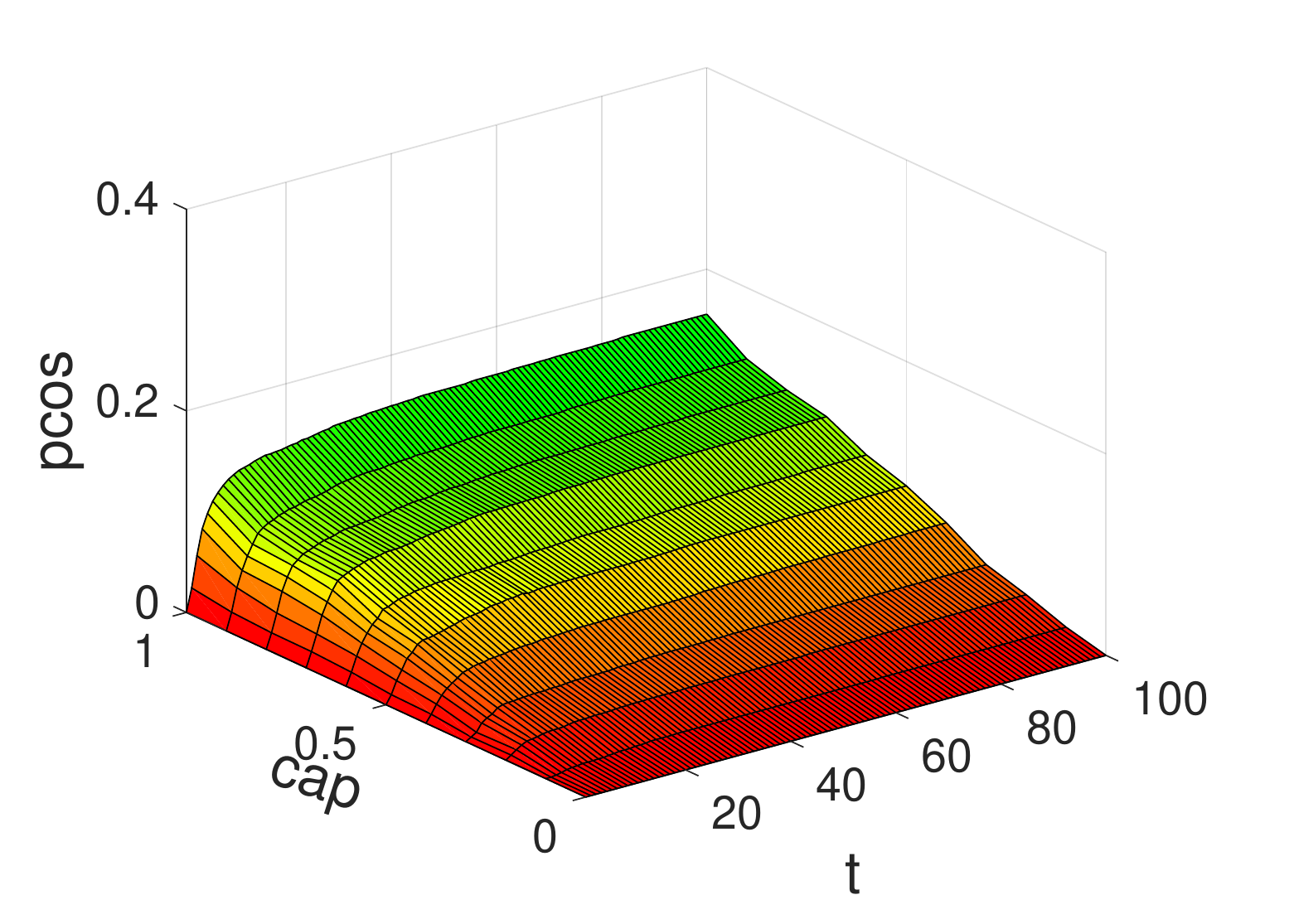}}
\subfigure[Configuration $C_0$, $\zeta=0.1$]{\label{fig:subfig:e}\includegraphics[width=0.23\textwidth]{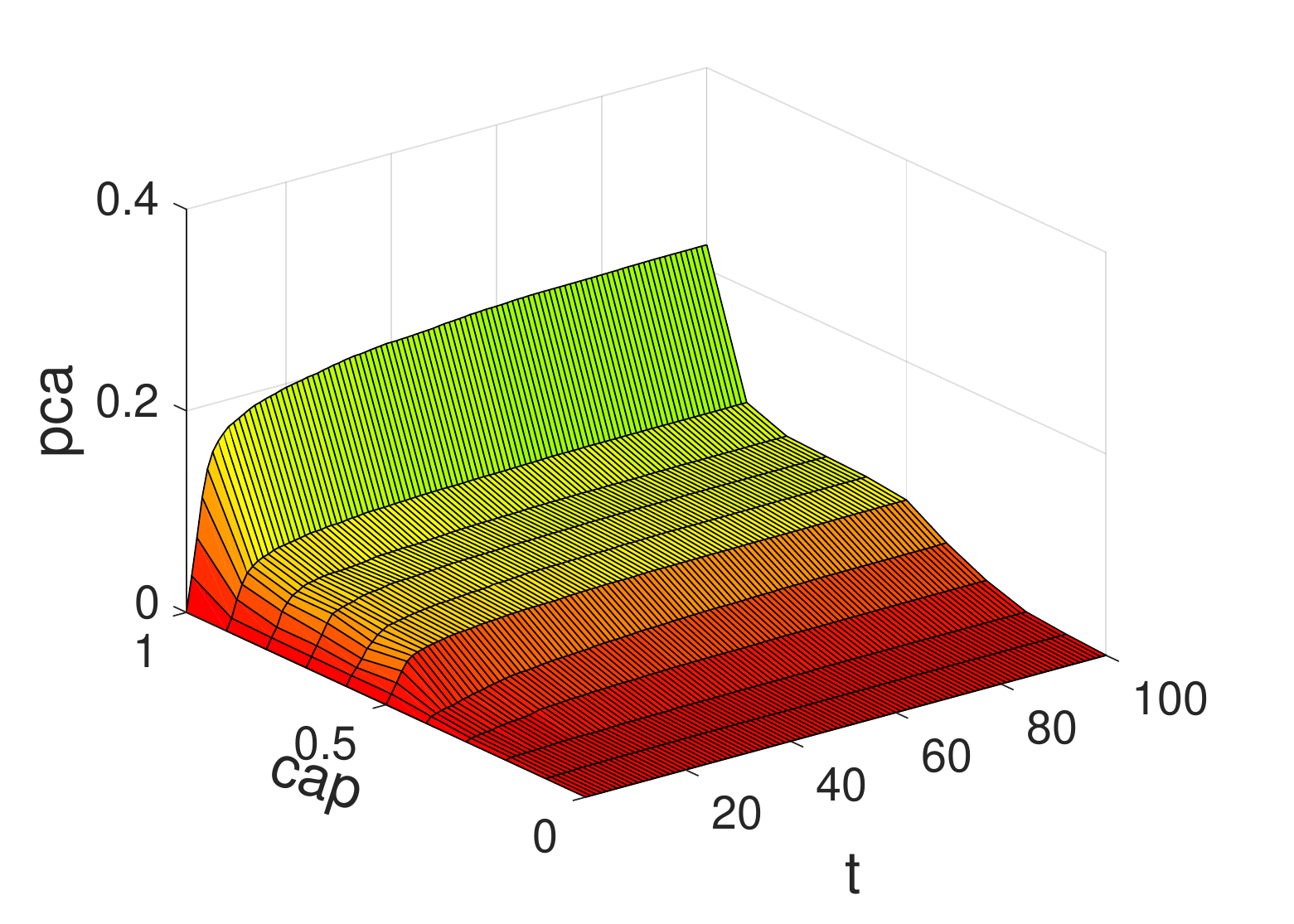}}
\subfigure[Configuration $C_1$, $\zeta=0.1$]{\label{fig:subfig:f}\includegraphics[width=0.23\textwidth]{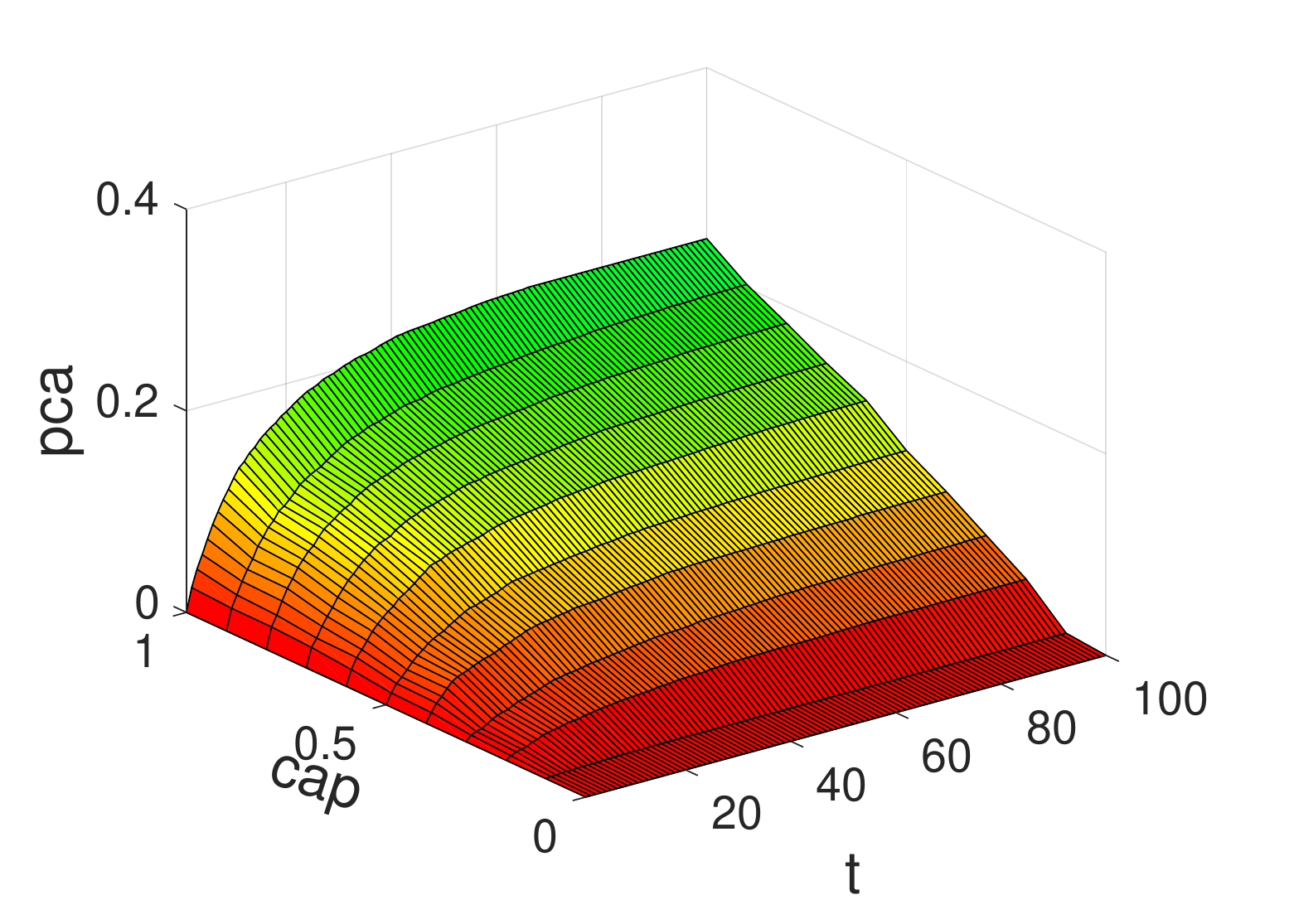}}
\caption{Plots of $\pca$ and $\pcos$ highlighting the smooth, rather than abrupt, decreases in security with increasing attacker capabilities.}
\label{diversity-effectiveness}
\vspace{-3mm}
\end{figure}

Fig. \ref{diversity-effectiveness} \comb{plots} $\pca$ and $\pcos$ with respect to attacker capability $\capa$ (the fraction of vulnerabilities that can be exploited by the attacker) and time. Figs. \ref{fig:subfig:a} and \ref{fig:subfig:c} show that monoculture, $C_0$, can lead to sudden ``jumps'' in terms of compromised applications and operating systems, namely that a single vulnerability can cause the compromise of many software. However, Figs. \ref{fig:subfig:b} and \ref{fig:subfig:d} show that the enforcement of network diversity, $C_1$, does not suffer from this problem. This shows that diversity can make the damage increases smoothly rather than abruptly, or make security degrades gradually rather than abruptly, with respect to increasing attack capabilities.

Suppose $\zeta$ is fixed, meaning that the security quality of independent implementations (in terms of their probabilities of being vulnerable) is the same. Suppose the attacker capability of the attacker $\capa$, namely the number of exploits the attacker has, is proportional to the number of vulnerabilities. Figs. \ref{fig:subfig:a} and \ref{fig:subfig:b} show that $\pca(100)$ with respect to $C_0$ and $\pca(100)$ with respect to $C_1$ are almost the same, meaning that enforcing software diversity does not lead to better security. This is because increasing $N$ also increasing $N\times \zeta \times \capa$ proportionally for fixed $\zeta$ and $\capa$. This example highlights when diversity neither increases nor decreases security.

By comparing Figs. \ref{fig:subfig:a} and \ref{fig:subfig:f}, we observe that diversity indeed leads to higher security when the diversified software have fewer vulnerabilities than the monoculture case. Indeed, the security resulting from diversity is almost proportional to the improvement in software security quality, namely the improvement in terms of reducing $\zeta$ (the probability that a software is vulnerable) from 0.2 to 0.1. This example highlights when diversity leads to higher security.

When comparing Figs. \ref{fig:subfig:b} and \ref{fig:subfig:e}, we observe that diversity actually can lead to lower security when the security quality of diversified implementations is poor. Indeed, the security resulting from using diversified low-quality software is almost proportional to the security quality of the software. For example, considering $\capa=1$ and $\pca(100)$, the damage of using low-quality diversified software ($\zeta=0.2$) is 1.5 times of the damage of using high-quality monoculture ($\zeta=0.1$). This highlights that diversity actually can lead to lower security when the diversified implementations are actually more vulnerable; this is possible because independently implementing multiple versions would incur a higher cost.
\begin{insight}
Network diversity can lead to gradually (rather than abruptly) increasing damages when the attacker gets more powerful. However, the security effectiveness of network diversity largely depends on the security quality of the diversified implementations, meaning that diversity can increase, make no difference, or decrease security, depending on the relative quality between the diversified implementations and the monoculture implementation.
\end{insight}

\subsubsection{Is the resulting security linear to the degree of diversity $N$ when attack capability is fixed?}
Suppose the attacker has a fixed number of exploits, say $|X|=2$ for each software.
Because enforcing diversity at all layers ($C_1$) may lead to a higher security, we now investigate
the effect of $N$. The parameters are: $\gamma = 0.2$, $\alpha = 0.2$, $\vartheta(\vul) = 0.8$, $\tau(\vul) = 0.05$, $\zeta=0.2$, $\capa=0.2$, $\omega=0.2$, NIPS= $tight$, HIPS= $tight$, and $t=100$.
\begin{wrapfigure}{r}{0.23\textwidth}
\vspace{-2mm}
\centering
{\includegraphics[width=0.23\textwidth,height=0.2\textwidth]{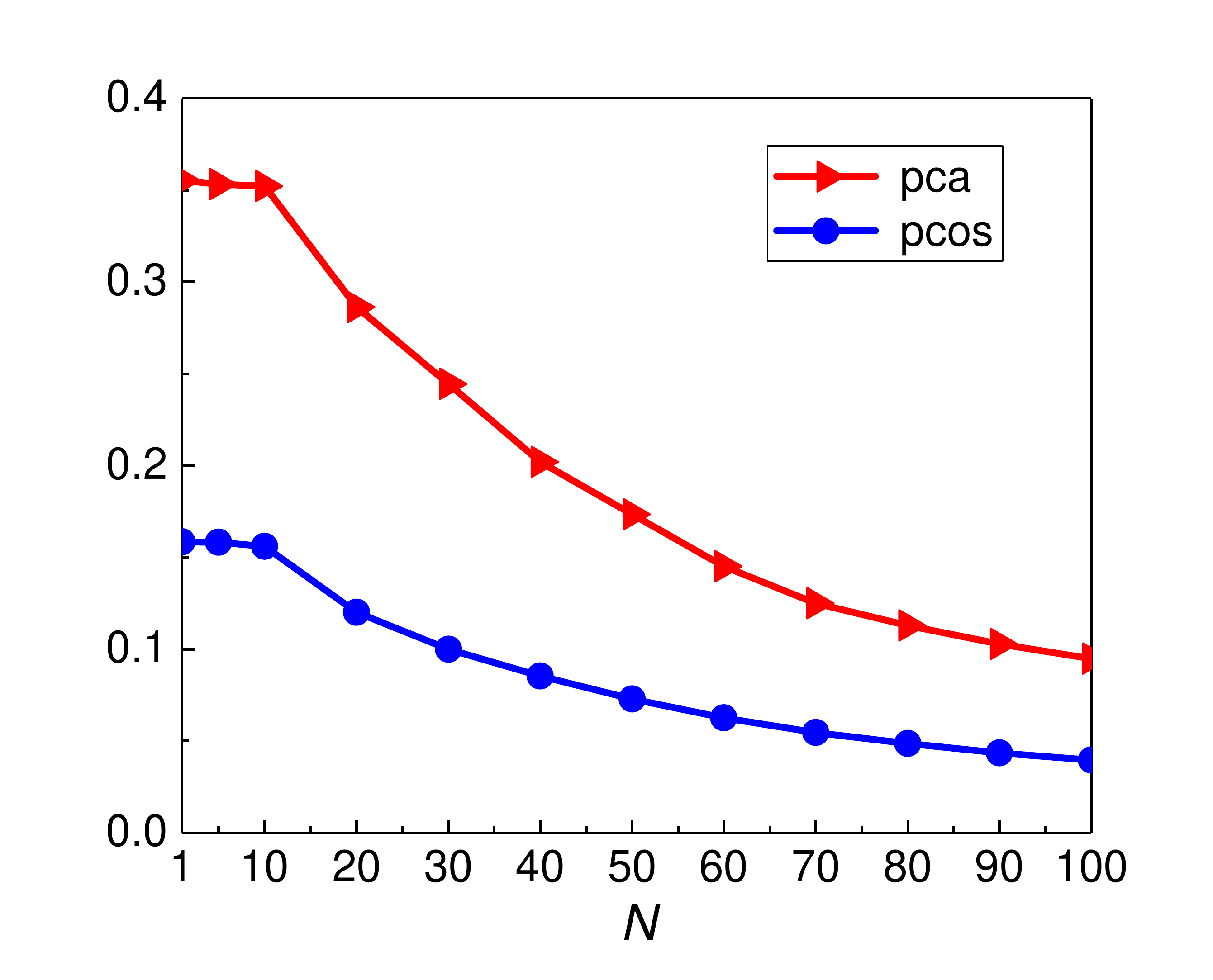}}
\caption{$\pca(100)$ and $\pcos(100)$.}
\vspace{-2mm}
\label{versions}
\end{wrapfigure}
Fig. \ref{versions} plots $\pca(100)$ and $\pcos(100)$ with varying $N$. We observe that when $N\leq10$, increasing $N$ does not lead to any significantly better security in terms of the two metrics, because in this case
each software has no more than $0.2\times 10=2$ vulnerabilities among all of the implementations, which can all be exploited.
When $N>10$, increasing $N$ does lead to better security because the attacker has only 2 exploits or exploit 2 vulnerabilities, even though there are, for example when $N=100$, $0.2\times 100=20$ vulnerabilities in the diversified implementations.
We further observe that security effectiveness, namely $1-\pca(100)$ and $1-\pcos(100)$, increases faster when $N\in [20,60]$ than $N\in[60,100]$.
This manifests a kind of ``diminishing return.'' Therefore, we have:

\begin{insight}
Given a fixed attack capability, increasing $N$ (diversity effort) leads to a higher security only when some vulnerabilities cannot be exploited by the attacker. Moreover, there appears to be a ``diminishing return'' in security effectiveness, highlighting the importance of considering cost-effectiveness in achieving network diversity.
\end{insight}

\noindent{\bf How to prioritize software diversity at the layers when diversity indeed improves security?}
Recall that configuration $C_0$ means monoculture, $C_1$ means enforcing diversity at all three layers, and $C_2,C_3,C_4$ respectively means enforcing diversity at the application, library, and operating system layer. In order to compare their effectiveness, we consider the following parameters: $\APP=\{$browser, email client, P2P, word processor$\}$, operating system is $\OS_1$, $N = 10$ (in the case of $C_1,\ldots,C_4$), $\gamma = 0.2$, $\alpha = 0.2$, $\vartheta(\vul) = 0.8$, $\tau(\vul) = 0.05$, $\zeta(v) = 0.2$ for $v$ not enforcing diversity, $\zeta(v) = 0.1$ for $v$ enforcing diversity, $\capa = 1.0$, $\omega = 0.2$, NIPS = $tight$, and HIPS = $tight$.

\begin{figure}[htbp!]
\vspace{-3mm}
\centering
\subfigure[$\pca(t)$]{\includegraphics[width=.23\textwidth]{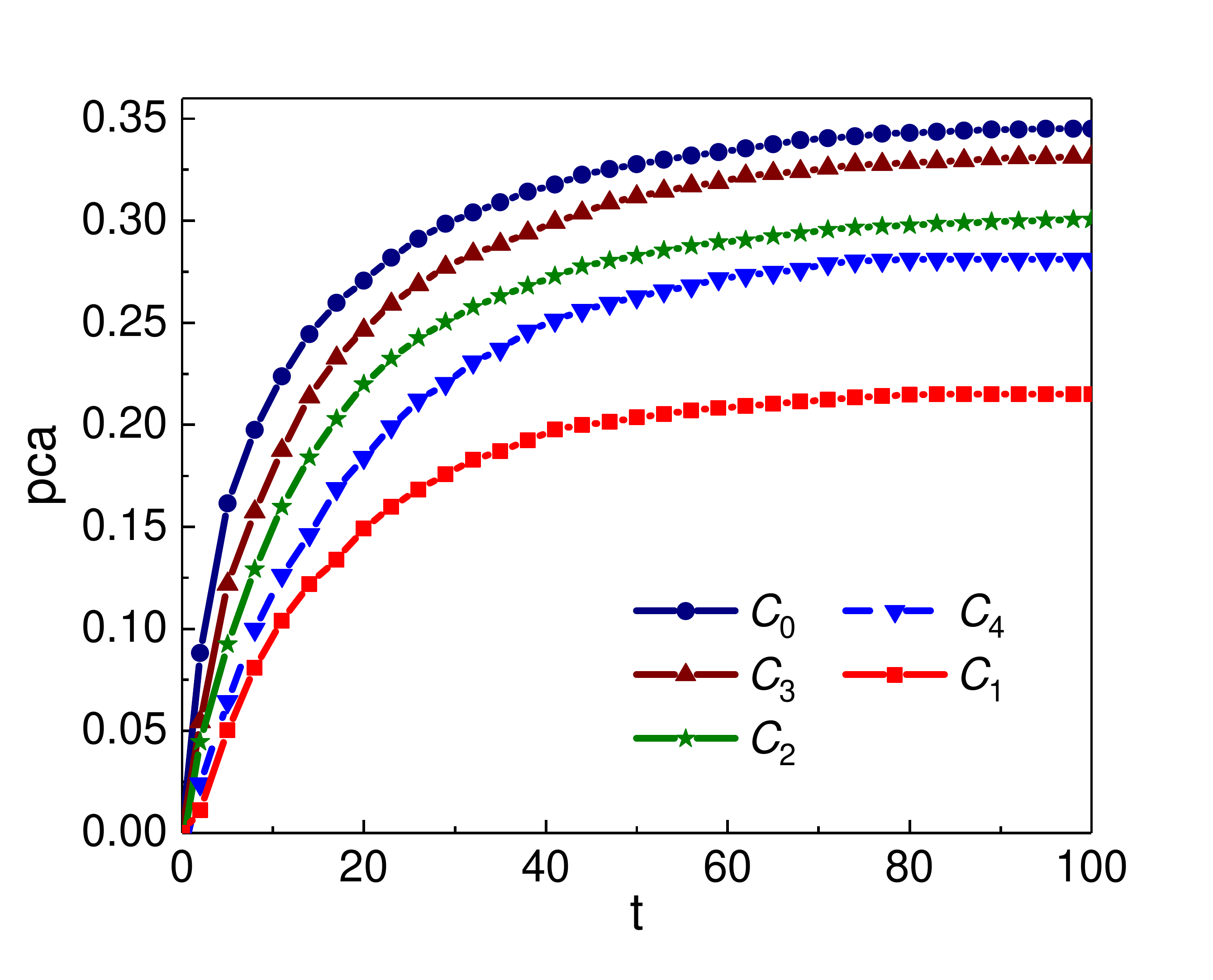}}
\subfigure[$\pcos(t)$]{\includegraphics[width=.23\textwidth]{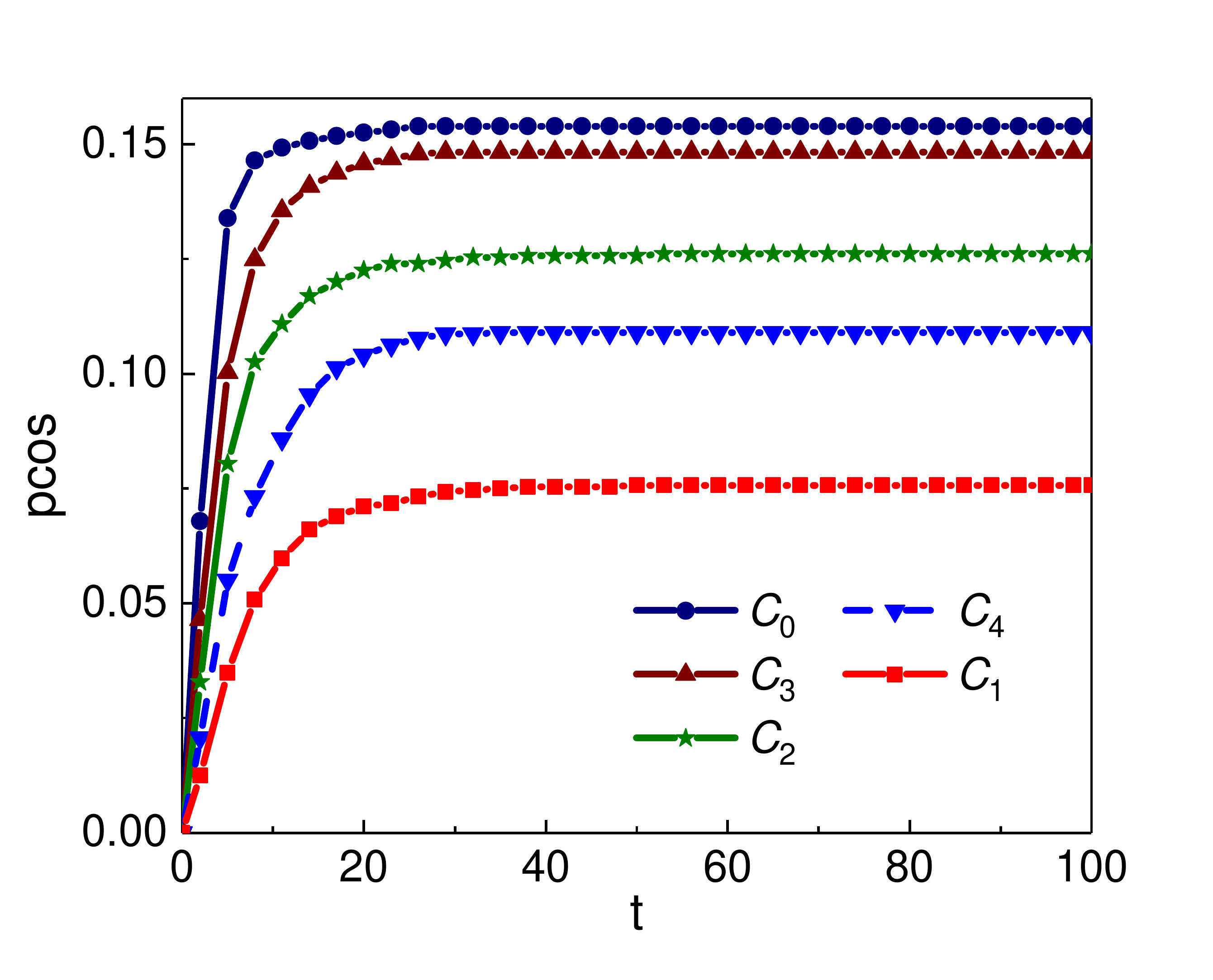}\label{multiple_layer_kernel}}
\vspace{-2mm}
\caption{$\pca(t)$ and $\pcos(t)$.}
\label{mul_layer_apps}
\vspace{-1mm}
\end{figure}

Fig. \ref{mul_layer_apps} plots  $\pca(t)$ and $\pcos(t)$ with different software stack configurations.
We observe that for the same configuration and parameters, $\pca(t)>\pcos(t)$ for any $t$, meaning that there are more compromised applications than compromised operating systems.
Moreover, we observe that both metrics $\pca(t)$ and $\pcos(t)$ show $C_0 \prec C_3 \prec C_2 \prec C_4 \prec C_1$, where $C_a \prec C_b$ means configuration $C_b$ leads to a higher security than $C_a$.
For example, $\pca(100)$ for $C_0$ is 1.64 times of $\pca(100)$ for $C_1$, meaning that enforcing diversity at all three layers can reduce 39\% of the damage when compared with the case of monoculture. This leads to:

\begin{insight}
When diversity can improve security, enforcing diversity at multiple layers leads to higher security than enforcing diversity at a single layer. Enforcing diversity at the operating system layer leads to higher security than enforcing diversity at the application layer, which leads to higher security than enforcing diversity at the library layer.
\end{insight}

\subsection{RQ3: Effectiveness of Hybrid Diversity}
In order to evaluate the security effectiveness of hybrid (i.e., natural and artificial) diversity, we consider a computer runs either $\OS_1$ or $\OS_2$. Suppose $\OS_1$ is diversified into $N_{\OS_1}$ implementations and $\OS_2$ is diversified into $N_{\OS_2}$ implementations. We consider three cases: (i) $N_{\OS_1}=0$ and $N_{\OS_2}=1$, meaning that all the computers run a monoculture operating system $\OS_2$; (ii) $N_{\OS_1}=1$ and $N_{\OS_2}=1$, meaning that each computer runs either $\OS_1$ or $\OS_2$ with probability 0.5, which corresponds to natural diversity; (iii) $N_{\OS_1}=10$ and $N_{\OS_2}=10$, meaning that each computer runs either $\OS_1$ or $\OS_2$ with probability 0.5, but both $\OS_1$ and $\OS_2$ have 10 independent implementations to choose, which corresponds to using hybrid (i.e., natural and artificial) diversity. The other parameters are: $\APP=\{$browser, email client, P2P, word processor$\}$, $C_1$ (every layer is artificially diversified), $\gamma = 0.2$ (the probability that attacks are not blocked by NIPS), $\alpha = 0.2$ (the probability that attacks that are not blocked by HIPS), $\vartheta(\vul) = 0.8$ (the probability that a vulnerability can be exploited remotely), $\tau(\vul) = 0.05$ (the probability that a vulnerability is zero-day), $\zeta(\OS_2)=0.4$, $\zeta(\OS_1)=0.2$, $\zeta(v)=0.2$ for $v\in V - \{\OS_1,\OS_2\}$, $\capa=1$ (the attacker can exploit every vulnerability, which corresponds to the worst-case scenario), $\omega=0.2$ (20\% of the nodes of $\Weapon$ are initially compromised), NIPS= $tight$, and HIPS= $tight$.

\begin{wrapfigure}{r}{0.23\textwidth}
\vspace{-2mm}
\centering
\includegraphics[width=0.23\textwidth,height=.2\textwidth]{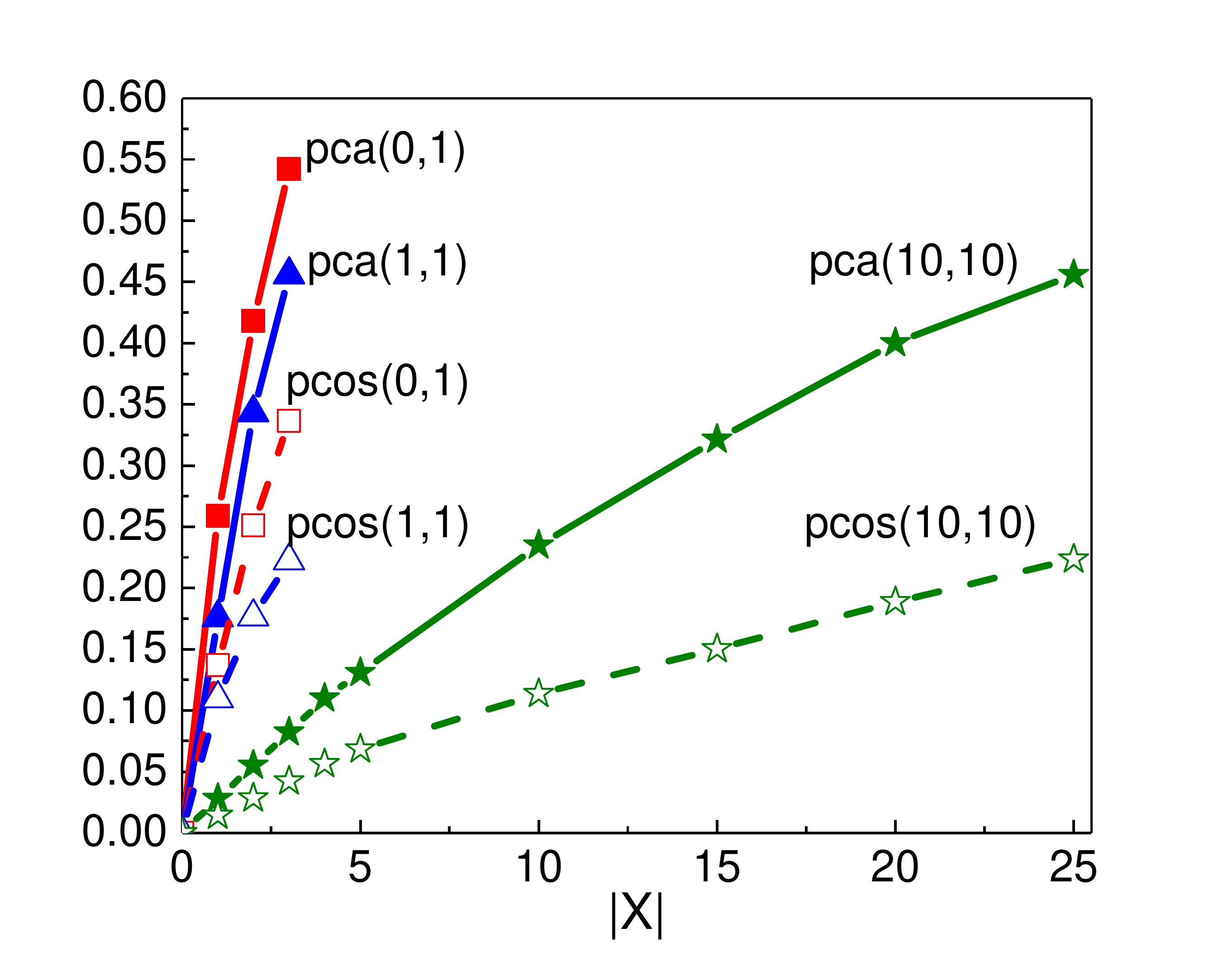}
\caption{Plot of $\pca(100)$ and $\pcos(100)$ with respect to $(N_{\OS_1},N_{\OS_2})$.}
\vspace{-2mm}
\label{figure:natural-diversity-single-vs-multiple-os}
\end{wrapfigure}

Fig. \ref{figure:natural-diversity-single-vs-multiple-os} plots $\pca(100)$ and $\pcos(100)$ with respect to $(N_{\OS_1},N_{\OS_2})$.
First, when comparing $\pca(0,1)$ and $\pca(1,1)$, we observe that natural diversity can lead to higher security once a higher quality of software is introduced, but the attacker's effort (the number of exploits $|\X|$) to compromise all of the vulnerable softwares remains the same.
Second, by comparing $\pca(1,1)$ and $\pca(10,10)$,
we observe that artificial diversity has no impact on security against a powerful attacker, but the attacker needs to pay almost 10 times the effort (or cost) to obtain exploits.
On the other hand, if the attacker's number of exploits, $|\X|$, is fixed,
artificial diversity can lead to a higher security. Third, by comparing $\pca(0,1)$ and $\pca(10,10)$, we observe that a comprehensive natural and artificial diversity not only can lead to a higher security, but also can substantially enhance the attacker's effort if diversified software implementations have higher security quality (i.e., less vulnerable). The same observations can be drawn from $\pcos(100)$.
This leads to:

\begin{insight}
It is beneficial to use both natural and artificial diversity (or ``diversifying the diversity methods''), assuming that the diversified implementations have at least the same quality.
\end{insight}

\subsection{RQ4: Quantifying Impact of Parameters}
\label{sec:regression-analysis}

As mentioned above, an important research goal is to obtain Eq. \eqref{eq:framework-equation}, namely
$m_i={\cal F}_i(G,A,B,C,D)$.
Due to the lack of real data,
we use, as a first step, {\em linear regression} to extract Eq. \eqref{eq:framework-equation} from the simulation data with respect to $t=100$ (i.e., the empirical steady state) and both natural and artificial diversities.
As a result, we can quantify the influence of each factor on $\pca$ and $\pcos$ and prioritize the factors that should be paid most attention in improving security.

For regression, we use the 14 explanatory variables $x_1,\ldots,x_{14}$ that are listed and explained in Table \ref{table:regression}, except for $x_5$ which is determined as follows:
$C_0,C_1,C_2,C_3,C_4$ respectively corresponds to $x_5=5,1,3,4,2$. This can be justified by the decreasing order of $\pcos(100)$ as shown in Fig. \ref{multiple_layer_kernel}, namely
$C_0,C_3,C_2,C_4,C_1$.
For the explanatory variables that are not defined over $[0,1]$, we first standardize them into $[0,1]$.
The data consists of 568 rows of these variables as well as the corresponding $\pca(100)$ and $\pcos(100)$.
Because 7 (out of the 14) variables are highly correlated with each other (e.g., the Pearson correlation coefficient between $x_5$ and $x_9$
is -0.3148, the coefficient between $x_6$ and $x_{10}$ is
0.4034, the coefficient between $x_{14}$ and $x_8$
is -0.45) and the amount of data is relatively small when compared with the number of variables (i.e., 14), we use the {\em partial least squares} method \cite{tobias1995introduction} to estimate the regression coefficients.
Since it should be the case that $\pca(100)=\pcos(100)=0$ when $x_1=\ldots=x_{14}=0$, the regression results are:
\begin{eqnarray}
\label{eq:regression}
\pca=\sum_{i=1}^{14}a_i x_i~~\text{and}~~\pcos=\sum_{i=1}^{14}b_i x_i,
\end{eqnarray}
where the $a_i$'s and $b_i$'s are respectively given in the 3rd and 4th column of Table \ref{table:regression}.
The cumulative R-square of $\pca(100)$ and $\pcos(100)$ in the fitted models is respectively 0.78 and 0.75, which means that the model fitting is accurate.

\begin{table}[!htbp]
\centering
\begin{tabular}{|p{1.2cm}<{\centering}|p{2.5cm}<{\centering}|c|c|}
\hline
\multirow{2}{*}{parameter} &\multirow{2}{*}{meaning}     &$\pca$          &$\pcos$  \\  \cline{3-3} \cline{4-4}
         &         &$a_i$          &$b_i$                  \\ \hline
$x_1$ & fraction of $\OS_1$ in network     & -0.075043802 & -0.087834035 \\ \hline
$x_2$ & \# apps running on a computer     & 0.088057385  & 0.101100359  \\ \hline
$x_3$ &fraction of $\browser_2$ in network   &0.030891731  & 0.021767401  \\ \hline
$x_4$ &$N$   & -0.135383668 & -0.132484815 \\ \hline
$x_5$ &(see text)      & 0.052477615  & 0.06748185  \\ \hline
$x_6$ &$\zeta\in [0,1]$   & 0.134057115  & 0.191437005  \\ \hline
$x_{7}$ &$\vartheta\in [0,1]$       & 0.329776965 & 0.195279456 \\ \hline
$x_{8}$ &$\tau\in [0,1]$      & 0.034082221  & -0.011789502 \\ \hline
$x_{9}$ &$\capa\in[0,1]$     & 0.730355334  & 0.593032     \\ \hline
$x_{10}$ &$\omega\in[0,1]$   & 0.170030442  & 0.087308213  \\ \hline
$x_{11}$ &$\alpha\in [0,1]$      & 0.050954156 & 0.05991876  \\ \hline
$x_{12}$ &$\gamma\in [0,1]$  &0.0001      &0.0001      \\ \hline
$x_{13}$ &NIPS=1 (i.e., {\em tight})   & -0.049658855  &- 0.035677035  \\ \hline
$x_{14}$ &HIPS=1  (i.e., {\em tight})        & -0.037981485 & -0.310823305 \\ \hline
\end{tabular}
\caption{Description of
the explanatory variables
and the dependent variables $\pca$ and $\pcos$.}
\label{table:regression}
\end{table}

Table \ref{table:regression} shows the following.
On one hand, the factors that have a significant influence on $\pca$ are $x_4, x_6,x_7,x_9,x_{10}$.
The most significant factor is $x_9$, namely the fraction of vulnerabilities that can be exploited by the attacker. This suggests that the most significant strategy is to reduce the fraction of vulnerabilities that can be exploited by the attacker.
On the other hand, the factors that have a significant influence on $\pcos$ are $x_2,x_4,x_6,x_7,x_9,x_{14}$.
The most significant factor is $x_{14}$, namely the tightness of HIPS. This suggests that the most significant strategy to improve operating system-layer security is to enforce tight HIPS, namely to preventing unauthorized applications, even if compromised, from waging privilege escalation attacks.

\begin{insight}
The most significant defense strategy is to reduce software vulnerabilities or prevent attackers from obtaining exploits.
The second most significant defense strategy is to enforce tight HIPS.
\end{insight}

\section{Related Work}
\label{sec:related-work}

Software diversity has been advocated for security purposes \cite{Geer2003,Stamp:2004:RM:971617.971650,zhang2001heterogeneous}. The most closely related prior work is perhaps \cite{ODonnellCCS2004}, which investigates how to configure diversified software implementations on computers. Their goal is to minimize the number of neighboring nodes that have the same software implementation, and/or maximize the number of subnetworks that run the same software. This means that \cite{ODonnellCCS2004} is {\em algorithmic} in nature. In contrast, we use the Cybersecurity Dynamics framework~\cite{XuCybersecurityDynamicsHotSoS2014} to model and analyze the security effectiveness of enforcing network diversity. When compared with \cite{ODonnellCCS2004}, our work can be characterized as follows: (i) we quantify the security effectiveness of enforcing network diversity, leading to insights that are not known until now; (ii) our model is fine-grained. Indeed, our model is finer grained than the numerous models in the Cybersecurity Dynamics framework (see, for example,~\cite{XuTDSC2011,XuTDSC12,XuTNSE2018}). This is because we explicitly model the caller-callee dependence relation between software components.

The idea of software diversity, especially N-version programming \cite{avizienis1985n,chen1978n}, was originally proposed to enhance fault tolerance under the assumption that software faults occur {\em independently} and {\em randomly}. Unfortunately, this assumption may not hold in general because programmers may make the same mistakes \cite{knight1986experimental,eckhardt1991experimental}, and because attacks are specifically geared towards software vulnerabilities (i.e., attacks are neither independent nor random). This means that the security value of enforcing software diversity must be re-examined in realistic threat models. To the best of our knowledge, the present study is the first effort aiming at systematically quantifying and characterizing the security effectiveness of software diversity {\em without} making the independence assumption between multiple implementations of the same software program. Indeed, we show that, in contrast to its fault-tolerance effectiveness, network diversity does {\em not} necessarily improve security when the diversified implementations possibly have the same security quality as the monoculture software implementation (i.e., containing the same amount of vulnerabilities).
The issue of {\em dependence} in the cybersecurity domain has been investigated in the Cybersecurity Dynamics framework, including the dependence between random variables \cite{XuInternetMath2012,XuQuantitativeSecurityHotSoS2014,XuInternetMath2015Dependence} and the dependence between cybersecurity time series data (instantiating stochastic processes)
\cite{XuVineCopula2015,peng2018modeling}. Indeed, dependence has been listed as one of the technical barriers that need to be adequately tackled for modeling and quantifying cybersecurity from a holistic perspective \cite{XuCybersecurityDynamicsHotSoS2014}.

Another specific method for achieving diversity is to use compiler techniques \cite{DBLP:journals/tdsc/HomescuJCBLF17,DBLP:conf/ccs/Franz15,DBLP:conf/ndss/CraneHBLF15,DBLP:journals/ieeesp/LarsenBF14}. In principle, the resulting diversified versions can be treated the same as the N-version programming. Moreover, our framework can accommodate a wide range of scenarios, from independence to dependence. We refer to \cite{DBLP:conf/sp/LarsenHBF14} for an outstanding systematization of knowledge in this diversification approach.
There are also proposals for runtime diversity,
including address space randomization \cite{bhatkar2003address,forrest1997building,etoh2000gcc,xu2003transparent}, instruction set randomization~\cite{kc2003countering,barrantes2003randomized},
and randomizing system calls \cite{chew2002mitigating}.
Effectiveness and weakness of these techniques have been analyzed in \cite{shacham2004effectiveness,sovarel2005s,DBLP:conf/ndss/RuddSBDHCLLDFSO17} from a building-block perspective rather than from the perspective of looking at a network as a whole.
Because these diversity techniques are complementary to software diversity, which is the focus of the present work, the present framework may be extended to investigate the security effectiveness of these techniques as well.
Moreover, researchers have proposed the notion of N-variant systems to achieve higher assurance in detecting attacks~\cite{cox2006n,holland2005architecture,knight1986experimental}.

Quantifying security is related to security metrics, for which there are three recent surveys \cite{Pendleton16,8017389,Noel2017} and some recent advancements are \cite{XuTIFSTrustworthiness2018,XuHotSoS2018Firewall}.
It is worth mentioning that our graph-theoretic framework is different from the framework of Attack Graphs
\cite{Phillips:1998:GSN:310889.310919,JhaOkland02},
because the former models the dynamics (i.e., time-dependent) and the latter is {\em combinatorial} in nature (i.e., time-independent).

\section{Limitations}
\label{sec:limitations}

We identify the following limitations of the theoretical framework and simulation study and leave them for future investigations.
First, the framework does not consider insider threats. Second, the framework focuses mainly on preventive defenses. Future research will include the investigation of a broader spectrum of defense mechanisms including reactive and adaptive defenses. Third, the simulation study assumes that firewalls cannot be compromised. This assumption can be eliminated, by accommodating the consequence of compromised firewalls (e.g., the network-based {\em tight} preventive defense enforced by a compromised firewall needs to become a {\em loose} preventive defense). Nevertheless, we already considered the security effectiveness of network-based {\em loose} preventive defense, which corresponds to the worst-case scenario in which all of the firewalls are compromised. It is worth mentioning that this issue is already resolved for host-based preventive defense, which is enforced by an operating system, because the compromise of an operating system already causes the compromise of the entire computer. Fourth, the $G=(V,E)$ used in the simulation study is heuristically generated, rather than derived from real-world network software stacks.

\section{Conclusion}
\label{sec:conclusion}

We proposed a theoretical framework to model network diversity, including a suite of security metrics for measuring attacker's effort, defender's effort, and security effectiveness of network diversity. We considered both natural and artificial diversities. We conducted simulation experiments to measure these metrics and draw insights from the experimental results. We characterized the conditions under which software diversity can lead to higher or lower security, or make no difference.
There are many problems for future research as the present study is just a first step to systematically quantify the security effectiveness of network diversity.

\smallskip

\noindent{\bf Acknowledgement}. We thank Lisa Ho for proofreading the paper.

\bibliographystyle{ieeetr} 

\end{document}